\documentclass[fleqn,usenatbib]{mnras}
\usepackage{newtxtext,newtxmath}

\usepackage[T1]{fontenc}

\DeclareRobustCommand{\VAN}[3]{#2}
\let\VANthebibliography\thebibliography
\def\thebibliography{\DeclareRobustCommand{\VAN}[3]{##3}\VANthebibliography}

\usepackage{graphicx,subcaption,ragged2e}
\usepackage{amsmath}	

\def\approxgt{\ifmmode \rlap{$>$}{}_{{}_{{}_{\textstyle\sim}}} \else$\rlap{$>$}{}_{{}_{{}_{\textstyle\sim}}}$\fi} 

\def\flx{erg cm$^{-2}$ s$^{-1}$}
\def\lum{erg s$^{-1}$}

\def\arcmin{\hbox{$^\prime$}}
\def\arcsec{\hbox{$^{\prime\prime}$}}
\def\approxlt{\ifmmode \rlap{$<$}{}_{{}_{{}_{\textstyle\sim}}} \else$\rlap{$<$}{}_{{}_{{}_{\textstyle\sim}}}$\fi}
\def\fxt{EP250207b}
\def\xmm{XMM-\emph{Newton}}
\def\ep{Einstein Probe}
\def\chan{\emph{Chandra}}

\title[\fxt~a compact binary merger?]{ \fxt~is not a collapsar fast X-ray transient. Is it due to a binary compact object merger?}
\author[Jonker et al.]{P. G. Jonker,$^{1,2}$
A. J. Levan,$^{1,3}$
Xing Liu,$^{4}$
Dong Xu,$^{4}$
Yuan Liu,$^{4}$
Xinpeng Xu,$^{4,5}$
An Li,$^{6}$
N. Sarin,$^{7,8}$
\newauthor N.R. Tanvir,$^{9}$
G.P. Lamb,$^{10}$ 
M.E. Ravasio,$^{1}$
J. S\'anchez-Sierras,$^{1}$
J.A. Quirola-V\'asquez,$^{1}$
B.C. Rayson,$^{9}$
\newauthor J.N.D. van Dalen,$^{1}$
D.B.~Malesani,$^{11,12,1}$
A.P.C. van Hoof,$^{1}$
F. E. Bauer,$^{13}$
J. Chac{\'{o}}n,$^{14}$
S.J. Smartt,$^{15,16}$
\newauthor A. Martin-Carrillo,$^{17}$
G. Corcoran,$^{17}$
L. Cotter,$^{17}$
A. Rossi,$^{18}$
F. Onori,$^{19}$
M. Fraser,$^{17}$
P.T. O'Brien,$^{9}$
\newauthor R.A.J. Eyles-Ferris,$^{9}$
J. Hjorth,$^{20}$
T.-W. Chen,$^{21}$
G. Leloudas,$^{22}$
L. Tomasella,$^{23}$
S. Schulze,$^{24}$
M. De Pasquale,$^{25}$
\newauthor 
F.~Carotenuto,$^{26}$
J.~Bright,$^{15, 27}$
Chenwei Wang, $^{28}$ 
Shaolin Xiong, $^{28}$ 
Jinpeng Zhang, $^{28}$ 
Wangchen Xue, $^{28}$ 
\newauthor 
Jiacong Liu, $^{28}$ 
Chengkui Li, $^{28}$ 
D. Mata S\'anchez,$^{29,30}$
M.A.P. Torres,$^{29,30}$
\\
$^{1}$Department of Astrophysics/IMAPP, Radboud University, 6525 AJ Nĳmegen, The Netherlands\\
$^{2}$SRON, Netherlands Institute for Space Research, Niels Bohrweg 4, 2333 CA Leiden, the Netherlands \\
$^{3}$Department of Physics, University of Warwick, Gibbet Hill Road, Coventry, CV4 7AL, UK \\
$^{4}$National Astronomical Observatories, Chinese Academy of Sciences, Beijing 100101, China\\
$^{5}$School of Astronomy and Space Science, University of Chinese Academy of Sciences, Beijing 100049, China\\
$^{6}$Beijing Normal University, No.19, Xinjiekouwai St, Haidian District, Beijing, 100875, P.R.China\\
$^{7}$Kavli Institute for Cosmology, University of Cambridge, Madingley Road, CB3 0HA, UK\\
$^{8}$Institute of Astronomy, University of Cambridge, Madingley Road, CB3 0HA, UK\\
$^{9}$School of Physics and Astronomy, University of Leicester, University Road, LE1 7RH, UK \\
$^{10}$Astrophysics Research Institute, Liverpool John Moores University, IC2 Liverpool Science Park, 146 Brownlow Hill, Liverpool, L3 5RF, UK\\
$^{11}$Cosmic Dawn Center (DAWN), Denmark\\
$^{12}$Niels Bohr Institute, University of Copenhagen, Jagtvej 128, 2200 Copenhagen N, Denmark\\
$^{13}$Instituto de Alta Investigaci{\'{o}}n, Universidad de Tarapac{\'{a}}, Casilla 7D, Arica, Chile\\
$^{14}$Instituto de Astrof{\'{\i}}sica, Facultad de F{\'{i}}sica, Pontificia Universidad Cat{\'{o}}lica de Chile, Campus San Joaquín, Av. Vicuña Mackenna 4860, Macul Santiago, Chile, 7820436\\
$^{15}$Department of Physics, University of Oxford, Keble Road, Oxford, OX1 3RH, UK \\
$^{16}$Astrophysics Research Centre, School of Mathematics and Physics, Queen’s University Belfast, BT7 1NN, UK\\
$^{17}$School of Physics and Centre for Space Research, University College Dublin, Belfield, Dublin 4, Ireland\\ 
$^{18}$Osservatorio di Astrofisica e Scienza dello Spazio, INAF, Via Piero Gobetti 93/3, Bologna, 40129, Italy\\
$^{19}$INAF-Osservatorio Astronomico d'Abruzzo, via M. Maggini snc, I-64100 Teramo, Italy \\
$^{20}$DARK, Niels Bohr Institute, University of Copenhagen, Jagtvej 155A, 2200 Copenhagen, Denmark\\
$^{21}$Graduate Institute of Astronomy, National Central University, 300 Jhongda Road, 32001 Jhongli, Taiwan\\
$^{22}$DTU Space, Department of Space Research and Space Technology, Technical University of Denmark, Elektrovej 327, 2800 Kgs. Lyngby, Denmark\\
$^{23}$Istituto Nazionale di Astrofisica INAF Osservatorio Astronomico di Padova via dell’Osservatorio 8, 36012 Asiago\\
$^{24}$Center for Interdisciplinary Exploration and Research in Astrophysics (CIERA), Northwestern University, 1800 Sherman Ave., Evanston, IL 60201, USA\\
$^{25}$University of Messina, MIFT Department, via F. S. D'Alcontres 31, Messina, 98166, Italy\\
$^{26}$ INAF-Osservatorio Astronomico di Roma, Via Frascati 33, I-00078, Monte Porzio Catone (RM), Italy\\
$^{27}$ Breakthrough Listen, Astrophysics, Department of Physics, The University of Oxford, Keble Road, Oxford OX1 3RH, UK\\
$^{28}$ State Key Laboratory of Particle Astrophysics, Institute of High Energy Physics, Chinese Academy of Sciences, 19B Yuquan Road, Beijing 100049, China \\
$^{29}$Instituto de Astrof\'isica de Canarias, IAC, E-38205, La Laguna, Tenerife, Spain \\
$^{30}$Departamento de Astrof\'isica, Univ. de La Laguna,   E-38206, La Laguna, Tenerife, Spain\\
}

\date{Accepted XXX. Received YYY; in original form ZZZ}

\pubyear{\the\year{2025}}

\begin{document}
\label{firstpage}
\pagerange{\pageref{firstpage}--\pageref{lastpage}}
\maketitle
 
\begin{abstract}
Fast X-ray Transients (FXTs) are short-lived extra-galactic X-ray
sources. Recent progress through multi-wavelength follow-up of
Einstein Probe discovered FXTs has shown that several are related to
collapsars, which can also produce $\gamma$-ray bursts (GRBs). In this
paper we investigate the nature of the FXT \fxt. The VLT/MUSE spectra
of a nearby (15.9~kpc in projection) lenticular galaxy reveal no signs
of recent star formation. If this galaxy is indeed the host, \fxt~lies
at a redshift of $z=0.082$, implying a peak observed absolute
magnitude for the optical counterpart of
$\mathrm{M_{r^\prime}=-14.5}$. At the time when supernovae (SNe) would
peak, it is substantially fainter than all SN types. These results are
inconsistent with a collapsar origin for \fxt. The properties favour a
binary compact object merger driven origin. The X-ray, optical and
radio observations are compared with predictions of several types of
extra-galactic transients, including afterglow and kilonova
models. The data can be fit with a slightly off-axis viewing angle
afterglow. However, the late-time ($\sim30$ day)
optical/NIR counterpart is too bright for the afterglow and also for
conventional kilonova models. This could be remedied if that late
emission is due to a globular cluster or the core of a (tidally
disrupted) dwarf galaxy. If confirmed, this would be the first case
where the multi-wavelength properties of an FXT are found to be
consistent with a compact object merger origin, increasing the
parallels between FXTs and GRBs. We finally discuss if the source could
originate in a higher redshift host galaxy.
\end{abstract}


\section{Introduction}

The first clear extra-galactic Fast X-ray Transients (FXTs) were detected serendipitously  (\citealt{Soderberg2008}; \citealt{Jonker2013}; \citealt{Bauer2017}). Systematic searches through 
\emph{Chandra} and XMM-\emph{Newton} archival data revealed $\approx$30 extra-galactic FXTs (\citealt{Glennie2015}; \citealt{Xue2019}; \citealt{Lin2019}; \citealt{Alp2020}; \citealt{Quirola2022}; \citealt{Eappachen2023}; \citealt{Quirola2023}). A small number of events later turned out to be caused by stellar flares from active stars in our Milky Way (e.g.~\citealt{Eappachen2024} reclassified the \xmm-discovered event XRT 140811 as a stellar flare). However, for the vast majority of identified extra-galactic FXTs, this scenario can be excluded. Nevertheless, without the detection of a contemporaneous counterpart at optical or near-infrared (NIR) wavelengths, their origin is difficult to determine. 

A small but critically important fraction of the observed sources in the transient sky are powered by the action of a compact central engine (black hole or highly magnetic neutron star). The prototype of these extreme events is the large population of $\gamma$-ray bursts (GRBs), whose nature has been the subject of intense study in the 50 years since their discovery. GRBs have typical durations spanning from a fraction of a second (short GRBs), to minutes (long GRBs; \citealt{kouveliotou93}), with only a tiny minority having durations up to a few hours (so-called ultra-long GRBs; \citealt{levan14}). They originate during the final moment of a star, either via the collapse of a massive stellar core \citep[e.g.~a collapsar;][]{hjorth03,stanek03} or the merger of two compact objects \citep[e.g.][]{tanvir13,berger13,abbott_bns}. Whereas initially long GRBs were exclusively associated with collapsars and short GRBs with binary compact object mergers, recent results show that the long-short -- collapsar-merger dichotomy is not strict (e.g.~\citealt{2022Natur.612..223R}).

Using the often arcsecond-precision knowledge of the FXT X-ray source position on the sky, deep searches reveal candidate host galaxies (e.g.~\citealt{Lin2022}; \citealt{Eappachen2024}; \citealt{Quirola2024b}). This in turn allows a (photometric) redshift to be derived, which sets the luminosity and energy scales involved, crudely constraining the nature of the FXT. However, it is only since the launch of the Einstein Probe (EP) satellite (\citealt{Yuan2022}; \citealt{2025SCPMA..6839501Y}) on Jan.~9, 2024 which is detecting about a 80 FXTs per year (depending on the signal-to-noise limit adopted) and announcing their discovery rapidly, that multi-wavelength counterparts to the FXTs have been discovered regularly (e.g.~\citealt{2024ApJ...969L..14G}; \citealt{2025GCN.39733....1Q}; \citealt{2025GCN.40260....1L}; \citealt{2025arXiv250421096A}).  The follow-up observations of counterparts led to the discovery of a broad-lined Ic supernova (SN) in the spectra and light curve of several EP-discovered FXTs, in particular, EP240414a (\citealt{vanDalen2025}; \citealt{Sun2025}; \citealt{2025ApJ...978L..21S}), EP250108a (\citealt{2025arXiv250408886E}; \citealt{2025arXiv250408889R}; \citealt{2025arXiv250417516S}), EP250304a (Cotter et al.~in prep.) and possibly EP241021a (\citealt{2025ApJ...985...21Z}; \citealt{2025arXiv250505444G}; \citealt{2025arXiv250314588B}; \citealt{2025arXiv250508781Y}; Quirola-Vasquez et al.~submitted). Similarly, for some 20-30\% of FXTs a (long) GRB is detected (e.g.~\citealt{Liu2024}; \citealt{Levan2024}; \citealt{2024ApJ...975L..27Y}; \citealt{2025arXiv250304306J}),  indicating that a significant fraction of the EP-discovered FXTs have a collapsar origin.

However, for an important fraction of \ep-discovered FXTs, no contemporaneous co-spatial burst of $\gamma$-ray emission is detected (\citealt{2024GCN.38435....1R, 2025GCN.39269....1R,2025GCN.40703....1R}, to list but a few) despite observations with sufficient sensitivity to detect such bursts for typical GRB spectral shapes. As several EP-discovered collapsar FXTs were also not detected in $\gamma$-rays, this is by no means evidence for a different nature than a collapsar origin for a significant fraction of FXTs. However, the FXT EP240408a does not seem to originate from a collapsar (\citealt{2025SCPMA..6819511Z}).

By analogy with the merger-driven and collapsar driven origins of GRBs, one could wonder if a fraction of FXTs can be linked to merger driven events, as has been suggested for many \chan-discovered FXTs, for instance on account of the plateau found in the X-ray light curve (e.g.~\citealt{2013ApJ...763L..22Z}; \citealt{2014MNRAS.439.3916M}; \citealt{2016ApJ...829...72C}; \citealt{2017ApJ...835....7S}; \citealt{2019ApJ...886..129S}; \citealt{Xue2019}; \citealt{Quirola2024}). With this in mind, we report here on the EP X-ray discovery of FXT \fxt~and our X-ray, optical, NIR, and radio follow-up observations.

Throughout this work, we assume the spatially-flat 6-parameter $\Lambda$CDM Planck cosmology (\citealt{2020A&A...641A...6P}) with $H_0=67.7$ km s$^{-1}$ Mpc$^{-1}$ and $\Omega_{m}$=0.31. We provide all magnitudes in the AB magnitude system. For NIR magnitudes calibrated to 2MASS, which is in the Vega system, we use the Vega to AB conversions  $J_{\mathrm{AB}} = J_{\mathrm{VEGA}} + 0.91$, $H_{\mathrm{}{AB}} = H_{\mathrm{VEGA}} + 1.39$, $K_{s,\mathrm{AB}} = K_{s,\mathrm{VEGA}} + 1.85$ (\citealt{2007AJ....133..734B}).

\section{Observations \& Results}

\subsection{Einstein Probe (EP) X-ray observations}
\subsubsection{EP -- Wide-field X-ray Telescope (EP-WXT) observations}
\label{wxt}
\begin{figure}
\centering
\includegraphics[width=1\columnwidth]{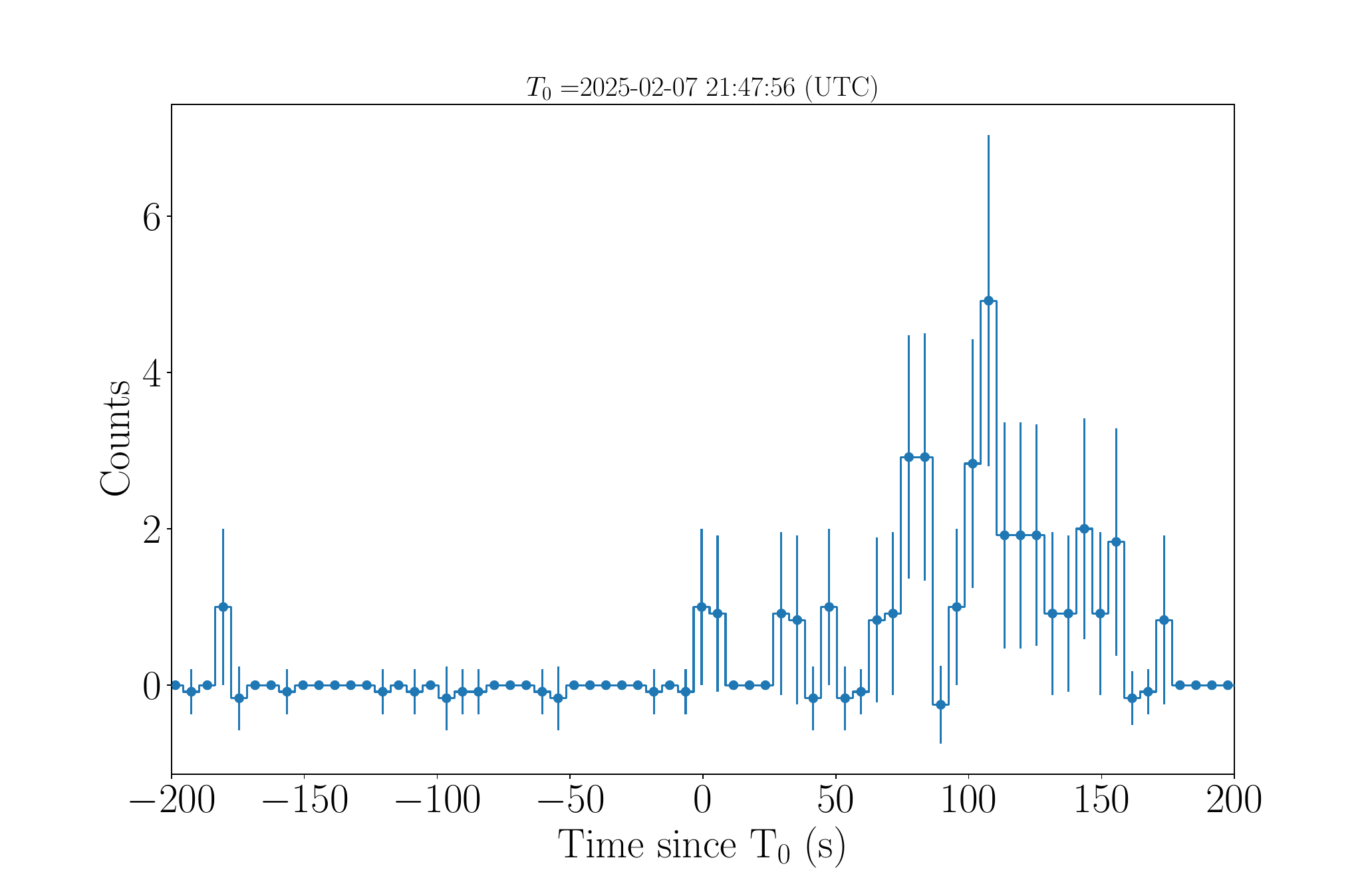}
\caption{The background subtracted \fxt~discovery light curve obtained by the EP-WXT instrument. Time is in seconds after T$_0$ and the bin size is 6~s. A period of $\approx$200~s before the FXT start is shown to assess the number of events at the source location before the FXT onset. }
\label{fig:lc-wxt}
\end{figure}

\begin{figure}
\centering
\includegraphics[width=0.99\columnwidth]{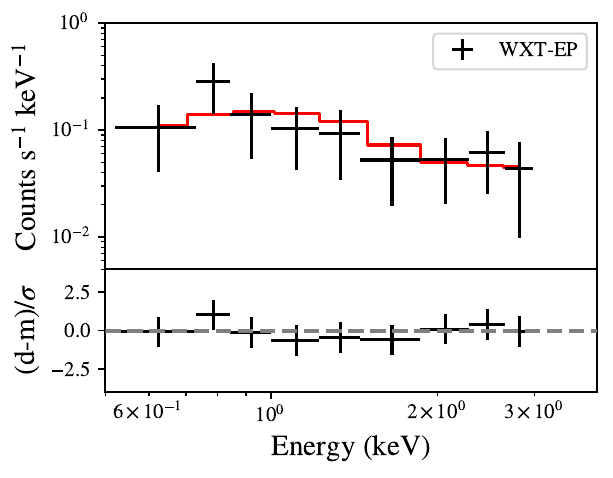}
\caption{{\it Top panel:} The \fxt~discovery spectrum (0.5--4 keV) obtained by the EP-WXT instrument averaged over 120~s. The best-fit power law model affected by Galactic extinction is shown. {\it Bottom panel:} The data, minus the best fit model, divided by the error in the data point is shown. No significant deviations with respect to the best-fit model are present. }
\label{fig:sp-wxt}
\end{figure}

A new FXT was discovered in EP-WXT observations on Feb.~7, 2025, at T$_0$=21:47:52.85 (UTC), which lasted more than 120~s and had a reported WXT position of Right Ascension (R.A.) = 167.495 deg (J2000), and Declination (Dec) = -7.906 deg (J2000) with a 90\% confidence uncertainty of 2.7\arcmin~ in radius 
\citep{2025GCN.39266....1Z}. After background subtraction, 27 source photons were detected in the 0.5-4 keV energy band. See Fig.~\ref{fig:lc-wxt} for the light curve. Requiring that each bin contains one or more photons and applying Poisson statistics in the fit, the average EP-WXT 0.5-4 keV spectrum during this period can be fitted well (Cash-statistics [\citealt{Cash1979}], 24.5 for 24 degrees of freedom [d.o.f.]) by an absorbed power law with a (fixed; \citealt{2016A&A...594A.116H}) line-of-sight Galactic equivalent hydrogen column density of 4$\times 10^{20}$~cm$^{-2}$ and a photon index of 0.5$\pm0.7$. The average unabsorbed 0.5-4 keV flux is $(6.5\pm3.6)\times10^{-10}$~\flx (90\% confidence level). See Fig.~\ref{fig:sp-wxt} for the best fit to the EP-WXT X-ray spectrum.  The EP X-ray data was processed using a data reduction pipeline and the calibration database (CALDB) specifically designed for WXT (Liu et al. in prep.). The CALDB incorporates results both from on-ground  and in-orbit calibration observations (\citealt{cheng-leia24}). We used {\sc XSPEC} version 12.15.0 for the fit (\citealt{1996ASPC..101...17A}). 

\begin{figure}
\centering
\includegraphics[width=0.95\columnwidth]{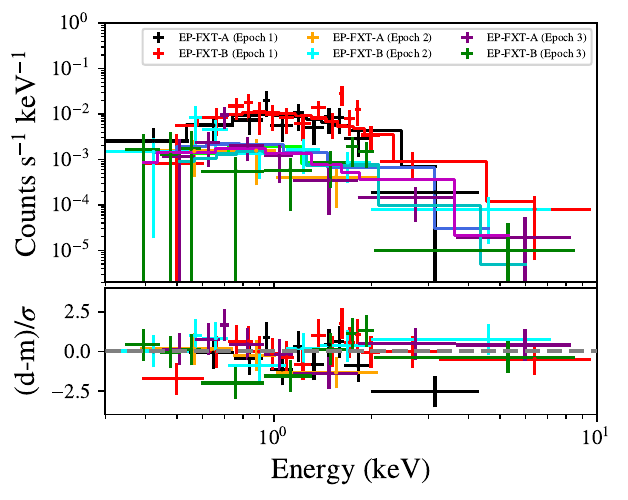}
\caption{{\it Top panel:} Shown is the EP-FXT spectrum for each of the three observations for each FXT telescope unit A and B separately. In addition, the best-fit power law model affected by Galactic extinction is shown using a matching-colour drawn line. Epoch 1 EP-FXT-A is shown in black and EP-FXT-B in red, while epoch 2 EP-FXT-A is shown in green and EP-FXT-B in blue, and finally, epoch 3 EP-FXT-A is shown in light blue and EP-FXT-B in purple. No large change in the source spectral shape is observed in between the three observing epochs while the flux decreased between the first and the second epoch. {\it Bottom panel:} The data, minus the best fit model, divided by the error in the data is shown for each of the EP-FXT-A and B unit telescope--detector system for each of the three observing epochs. No significant deviations with respect to the best-fit model are present. The colours represent the same data/epoch as in the {\it top panel.}}
\label{fig:sp-fxt}
\end{figure}

\subsubsection{EP-Follow-up X-ray Telescope (EP-FXT) observations} 
The EP-Follow-up X-ray Telescope instrument with acronym EP-FXT\footnote{Note that we always refer to the Einstein Probe Follow-up X-ray Telescope instrument as "EP-FXT" and to a Fast X-ray Transient as "FXT" to try to avoid confusion between the two.} consists of two telescopes each with its own detector, system A and B, and the detectors are sensitive over the 0.3--10 keV band. Three EP-FXT follow-up observations were performed. The first one started on Feb.~8, 2025, at 14:50:57 (UTC) with an exposure time of 3025~s (at $t=0.71$~d after the EP-WXT trigger). The second one started on Feb.~9, 2025 at 18:09:49 (UTC) with an exposure time of 5044~s (at $t=1.85$~d after the EP-WXT trigger). The results of these first two observations have also been announced in \cite{2025GCN.39266....1Z} with a best-fit source J2000 position of: (R.A.,Dec)~=~(167.5130, -7.8695)  with an uncertainty of 10\arcsec~(radius, 90\% confidence level). A third observation started on Feb.~10, 2025 at 13:16:43 (UTC) with an exposure time of 9045~s (at $t=2.65$~d after the EP-WXT trigger). 

The spectra of these EP-FXT observations (see Fig.~\ref{fig:sp-fxt}) have been fit simultaneously using an absorbed power law with a Galactic equivalent hydrogen column density fixed to the mean line-of-sight value provided in \cite{2016A&A...594A.116H}
of 4$\times 10^{20}$ cm$^{-2}$ and photon indices of 1.6$\pm0.4$, 2.3$\pm$0.9, and 2.3$\pm$0.9, for the first, second, and third observations, respectively. In the fit, the EP-FXT-A and EP-FXT-B detectors have their own response matrices. The fit has a Cash-statistic of 156 for 147 d.o.f. The unabsorbed 0.3--10 keV fluxes are $(3.3^{+1.4}_{-0.9}) \times 10^{-13}$, $(4^{+3}_{-1}) \times 10^{-14}$, $(3^{+2}_{-1}) \times 10^{-14}$~\flx~(90\% confidence level), respectively. No significant change in the power law spectrum is observed over the three observing epochs while the flux decreased, at least between the first and second epochs. 
We also investigated the data of the EP-FXT-A and EP-FXT-B detectors for each of the three epochs to search for variability (e.g.~flares), but none were found. Note that each detector only detected between 17--61 counts during these observations.

\begin{figure}
\centering
\includegraphics[width=0.95\columnwidth]{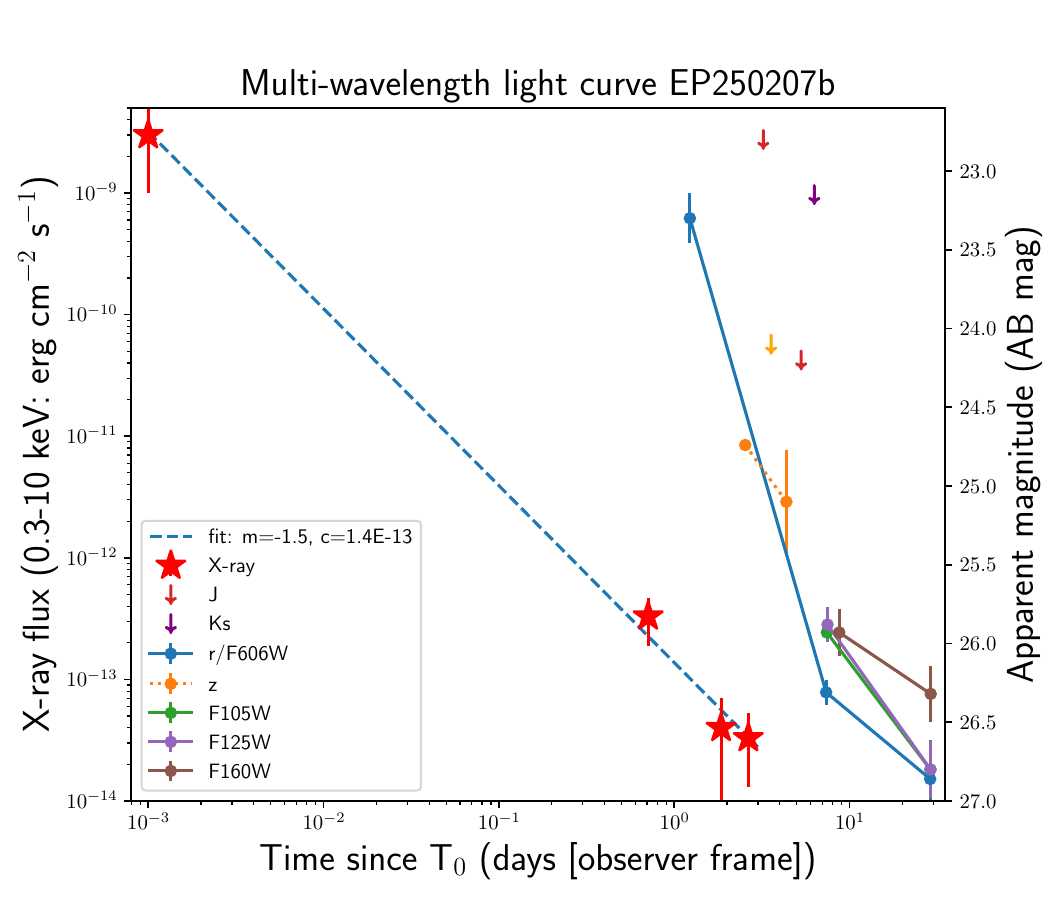}
\caption{The EP 0.3--10 keV X-ray light curve of \fxt. The first data point is the EP-WXT average flux and the next three data points are from the EP-FXT observations. The blue dashed line indicates the best fit power law function of $F_X=C\times (\frac{t}{1~\mathrm{d}})^m$. The location of the first data point is taken to be at $t=10^{-3}$~d ($\approx86$~s), reflecting that the first data point is an average over $\approx150$~s. The right hand y-axis shows the observed optical and NIR magnitudes. Clear fading is detected in the different filters (see Sections~\ref{groundbased} \& \ref{hst} ).}
\label{fig:lc}
\end{figure}

We converted the 0.5-4 keV EP-WXT (unabsorbed) flux to a 0.3--10 keV (unabsorbed) flux to facilitate comparison with the EP-FXT unabsorbed flux using W3PIMMS taking the best-fit absorbed power law model to the EP-WXT spectrum (see Section~\ref{wxt}) as input. We obtain an EP-WXT unabsorbed 0.3--10 keV flux of $(3^{+2}_{-1.4})\times 10^{-9}$~\flx. We show the X-ray light curve combining the EP-WXT and EP-FXT measurement in Fig~\ref{fig:lc}. Over-plotted is the best-fit power law decay function with $F_X=C\times (\frac{t}{1~\mathrm{d}})^m$, the best-fit power law index $m=-1.5$ and $C=1.4\times 10^{-13}$~\flx.

\subsection{GECAM limits on gamma-ray emission}
The Gravitational wave high-energy Electromagnetic Counterpart All-sky Monitor (GECAM) is a constellation of four gamma-ray all-sky monitors, including GECAM-A/B (\citealt{2022RDTM....6...12L}), GECAM-C (\citealt{2023NIMPA105668586Z}), and GECAM-D (\citealt{2024ExA....57...26W}). Throughout the burst duration of \fxt, only GECAM-B was continuously collecting data with a good coverage of the location of \fxt. However, no significant signal was detected neither in-flight (\citealt{2024RAA....24j4002Z}) nor on-ground (\citealt{2025SCPMA..6839511C}). We performed a targeted search (\citealt{2021MNRAS.508.3910C}) for the detection of a source using GECAM-B data from 2025-02-07 21:47:36 to 2025-02-07 21:56:16 (UTC). No significant source is found. We calculated an upper limit to the detection of a source  considering three typical GRB spectral models (i.e. soft, normal and hard Band functions (following e.g.~\citealt{2025SCPMA..6871011Z}) and three timescales (0.1~s, 1~s, 10~s). The 3 sigma upper limits (\citealt{2025ApJ...987L..38Z}) on the GRB flux (10-1000 keV) vary between  $\approx (1-10)\times10^{-7}$\flx~for the longest hard to the shortest soft assumed GRBs.
We also checked for a potential $\gamma$-ray counterpart of EP250207b in Fermi/GBM. Unfortunately, the location of the source was Earth-occulted for Fermi/GBM for the entire duration of the event, so no simultaneous Fermi/GBM upper limit can be reported. Finally, we checked whether the EP250207b location was observed by Swift/BAT. While the location of the source was not Earth-occulted for Swift/BAT, the position fell outside the coded-mask field of view. As a result we do not report a Swift/BAT upper limit.

\subsection{Optical and near-infrared ground based observations}
\label{groundbased}
The photometry reported below is uncorrected for extinction, which taking the $\mathrm{N_H}=4\times 10^{20}$~cm$^{-2}$ from the X-ray spectral fits, would correspond to an $\mathrm{A_V}=0.18$~mag following \citet{2009MNRAS.400.2050G}. This is slightly higher than the Galactic $\mathrm{A_V}=0.14$~mag derived from the dust reddening in SDSS stars (\citealt{2011ApJ...737..103S}, assuming $\mathrm{\frac{A_V}{E(B-V)}} = 3.1 $). Given the intrinsic scatter in the relation, this is likely consistent with the Galactic $\mathrm{A_V}$ value (cf.~\citealt{2017MNRAS.471.3494Z}).

\subsubsection{NOT/ALFOSC + NOT/NOTCam}

\begin{figure}
\centering
\includegraphics[width=0.95\columnwidth]{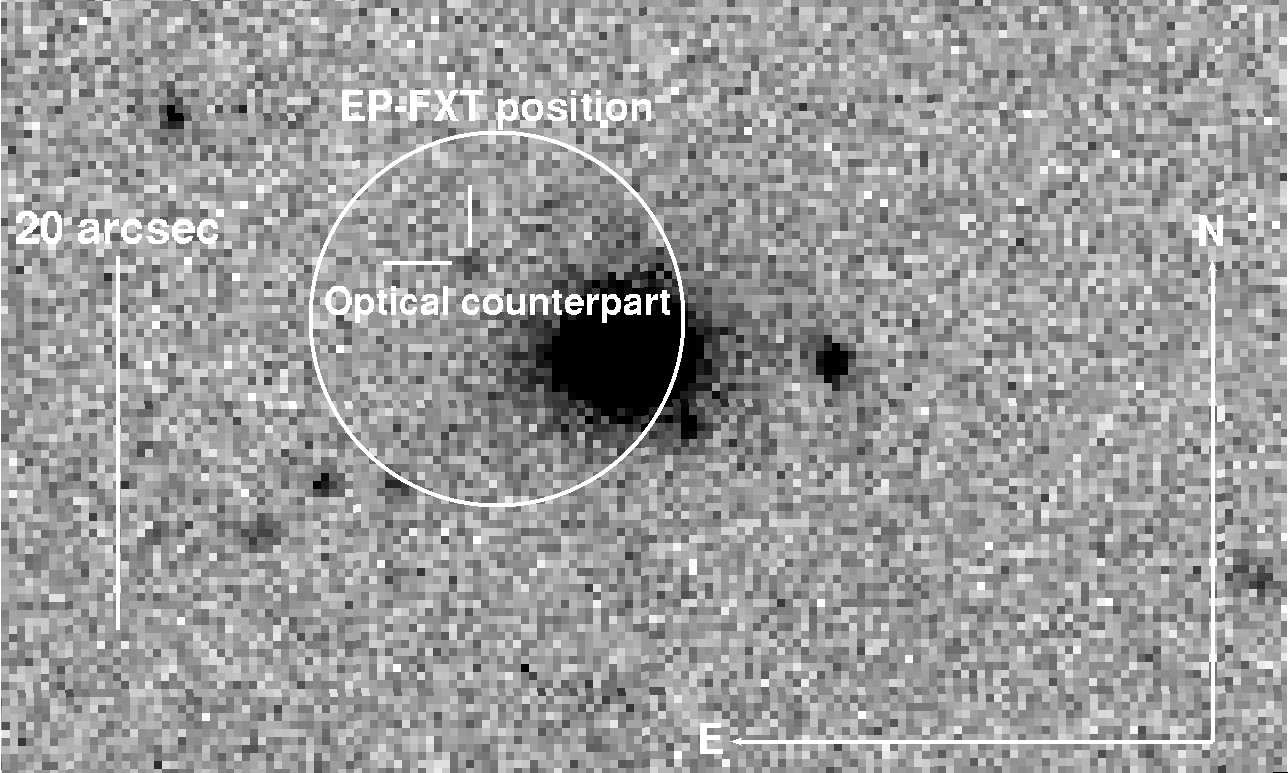}
\caption{The NOT/ALFOSC discovery $r^\prime$-band image of the optical counterpart to the FXT \fxt~combining the two best-seeing images of the four. The white circle indicates the EP-FXT source localization uncertainty region (\citealt{2025GCN.39266....1Z}). The new faint optical source ($r^\prime=23.3\pm0.16$~mag) is indicated by the tick marks. It lies in projection close to the galaxy WISEA~J111002.65$-$075211.9.}
\label{fig:not-cp}
\end{figure}

The field of \fxt\,was observed twice using the Nordic Optical Telescope (NOT). Initially, the ALFOSC instrument was used to obtain 4 exposures of 200~s in the $r^\prime$-filter. The mid-time of these exposures was 2025-02-09 03:16:13 UTC, i.e.~1.23~d after the EP-WXT start time of \fxt. After standard bias subtraction and correction for flatfield, a source was discovered that is not present in deeper Legacy Survey images (\citealt{dey2019}) of the field. We subtract the Legacy $r^\prime$-band image from the new NOT image to remove background light at the position of the counterpart due to the nearby (candidate host) galaxy using the ZOGY algorithm \citep{Zackay2016} implemented in PyZOGY \citep{Guevel2021}. From the subtracted image, we obtain a magnitude of $r^\prime=23.3\pm0.16$ for the new source (calibrated against Pan-STARRS). This is broadly consistent with the magnitude of $r^\prime\approx$23.7 quoted in the original work  reporting on this data \citep{2025GCN.39300....1L}. In Fig.~\ref{fig:not-cp} we show the NOT counterpart $r^\prime$-band discovery image.

At a mid-time of 2025-02-11 03:07:13 UTC, i.e.~3.22~d after the onset of the FXT, 30 exposures of 60~s each were obtained in the $J$-band filter using the NOTCam detector. No source was detected in the combined image at the location of the candidate optical counterpart to the FXT down to $J_{\mathrm{AB}}>$22.8 mag (3~$\sigma$).

In order to determine the best position of the source, we combined the first and last of our four ALFOSC images as they were taken under the best seeing conditions. The best fit position of the transient source has a R.A.~(J2000) = 11:10:03.22 and Dec.~(J2000) = -07:52:07.25 with an estimated uncertainty of $\approx$0.5\arcsec~(68 \% confidence), which falls well inside the EP-FXT error region of \fxt~ (see Fig.~\ref{fig:not-cp}).

\subsubsection{Gemini North and South Multi-Object Spectrograph observations (GMOS)}

The field of the FXT was observed using the GMOS instrument, on both the Gemini-South (GS) telescope located at Cerro Pachon, Chile, as well as the Gemini-North (GN) telescope located at Mauna Kea, Hawaii, U.S. GMOS was used in imaging mode during three epochs (two at GN at $\sim$2.54~d [6 exposures of 60~s each] and 3.57~d [5 exposures of 60~s each]), and one at GS 4.36~d after the discovery of \fxt\, [12 exposures of 60~s each] using programs GS-2024B-Q-131 and GS-2024B-FT-112 (PI Bauer). Two GMOS observations were executed using the $z^\prime$-filter and one using the $g^\prime$-filter. In the first GMOS observation, a faint source was detected at the position of the candidate optical counterpart with $z^\prime=24.7\pm0.2$~mag. The second $z^\prime$-filter observation yielded a non-detection with $z^\prime>24.1$~mag, while a deeper final GS/GMOS observation provided a faint detection with $z^\prime=25.1\pm0.3$~mag. The photometry was calibrated against Pan-STARRS.

Furthermore, we used GS/FLAMINGOS~2 (F2) at two different epochs to observe the field of \fxt\, in the $J$ and $K_s$-filters. The $J$-band observations were obtained on 2025-02-13, the first and last exposures started at 04:46:41 (UTC) and 05:12:35 (UTC), respectively. We combined 27 exposures of 40~s to search for emission from the counterpart. No source was detected at the position of the counterpart, with a 3~$\sigma$ upper limit of $J_{\mathrm{AB}}>$24.2~mag using the Photometry Sans Frustration python tool (\citealt{2023ApJ...954L..28N}). In addition, on 2025-02-14 starting at 04:39:23 (UTC) 90 exposures of 15~s each were obtained using the $K_s$-filter. The last exposure started at 05:33:11 (UTC). We derive an upper limit at the source position of $K_{s,\mathrm{AB}}>$23.15~mag (3~$\sigma$).

\subsection{Hubble Space Telescope observations}
\label{hst}
The field of the optical counterpart to \fxt\, was observed twice using the Wide Field Camera 3 (WFC3) onboard the \textit{Hubble Space Telescope} (HST) under program GO-17806 (PI Tanvir). Observations were obtained in four different wide-band filters and two detectors. During the first epoch 4$\times$505~s exposures were obtained totalling 2020~s in the F606W filter using the ultraviolet-visible (UVIS) detector and 4 exposures of 552.94~s were obtained totalling $\approx$2212~s in F105W, F125W and F160W each were obtained using the IR detector. All the second epoch observations had an identical set-up and exposure time as that of the first epoch. However, during the second epoch of F606W observations a cosmic ray hit very close to the transient's position was present in two of the four exposures. Therefore, we used only 2$\times$505~s totalling 1010~s of exposure in the F606W filter. 

\begin{figure*}
    \begin{subfigure}[t]{1.\linewidth}
        \includegraphics[width=\textwidth]{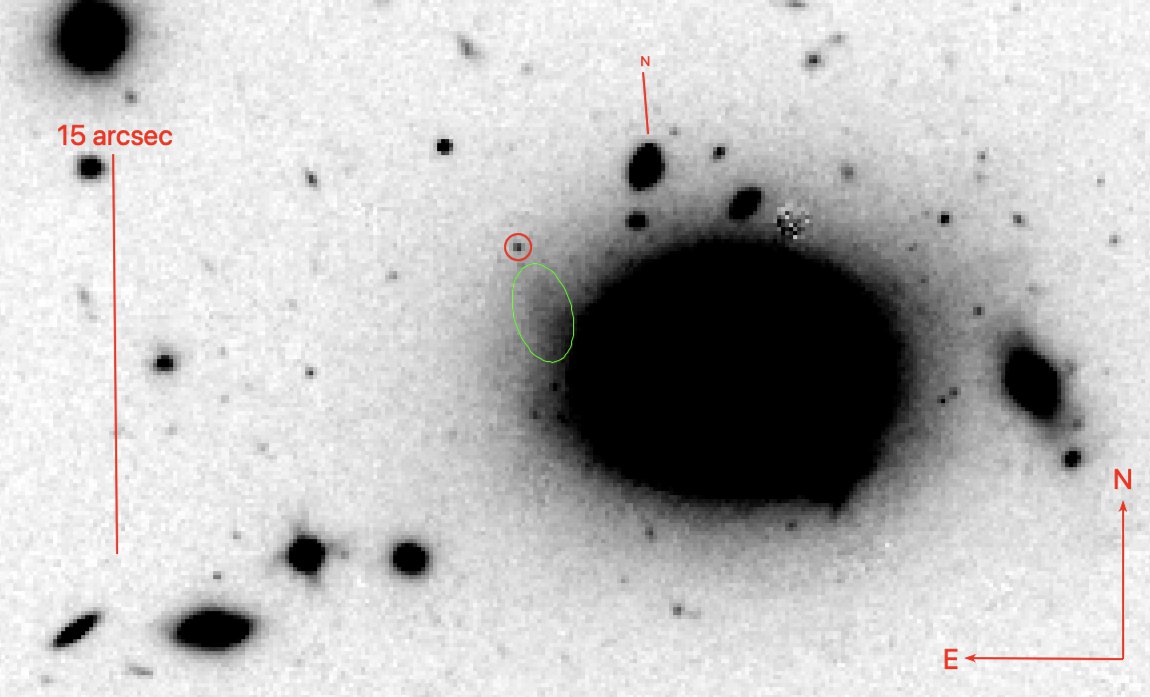}
    \end{subfigure}    \centering \hfill
    \begin{subfigure}[b]{1.\linewidth}
        \includegraphics[width=0.33\linewidth]{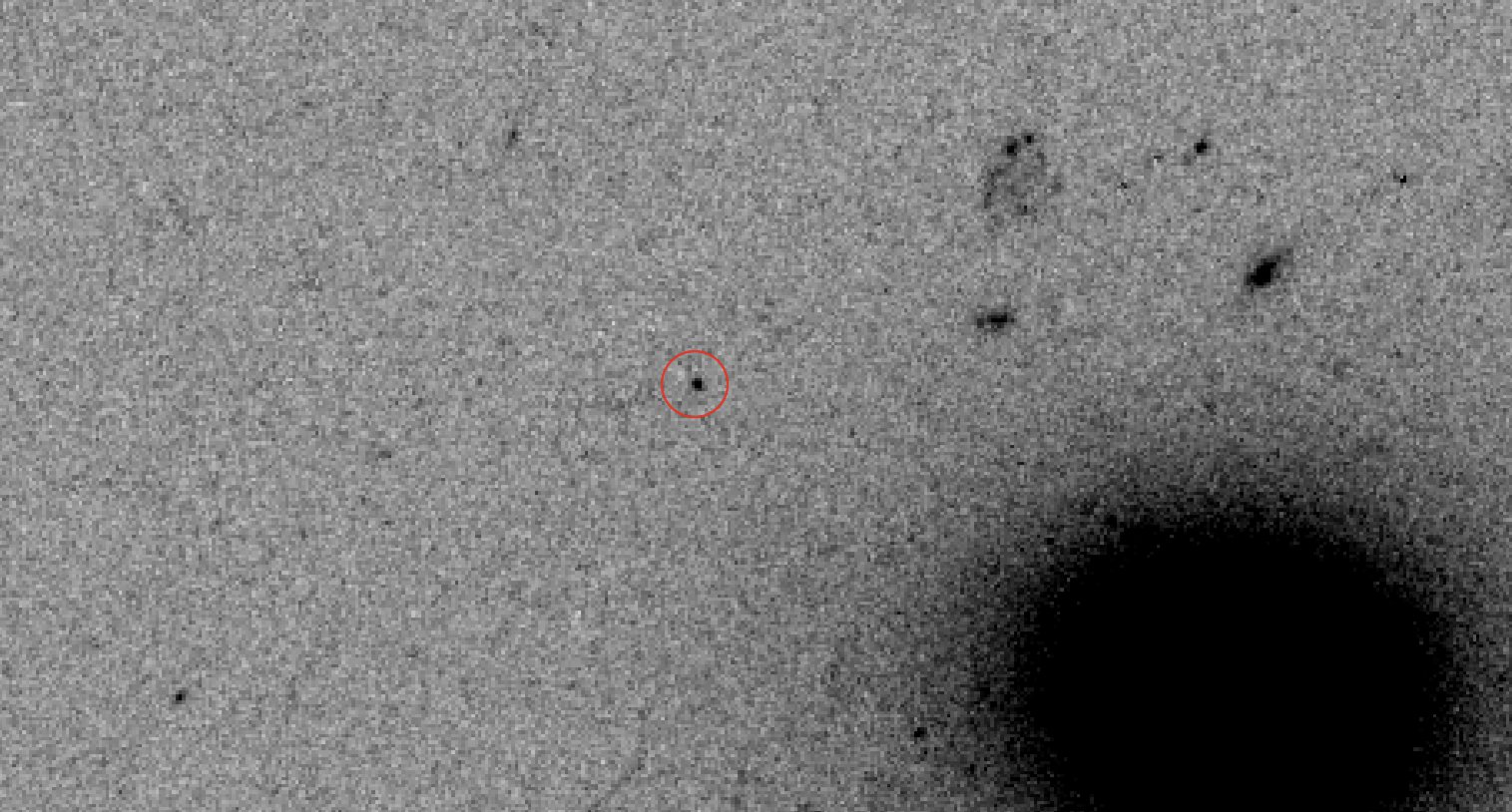}
        \includegraphics[width=0.33\linewidth]{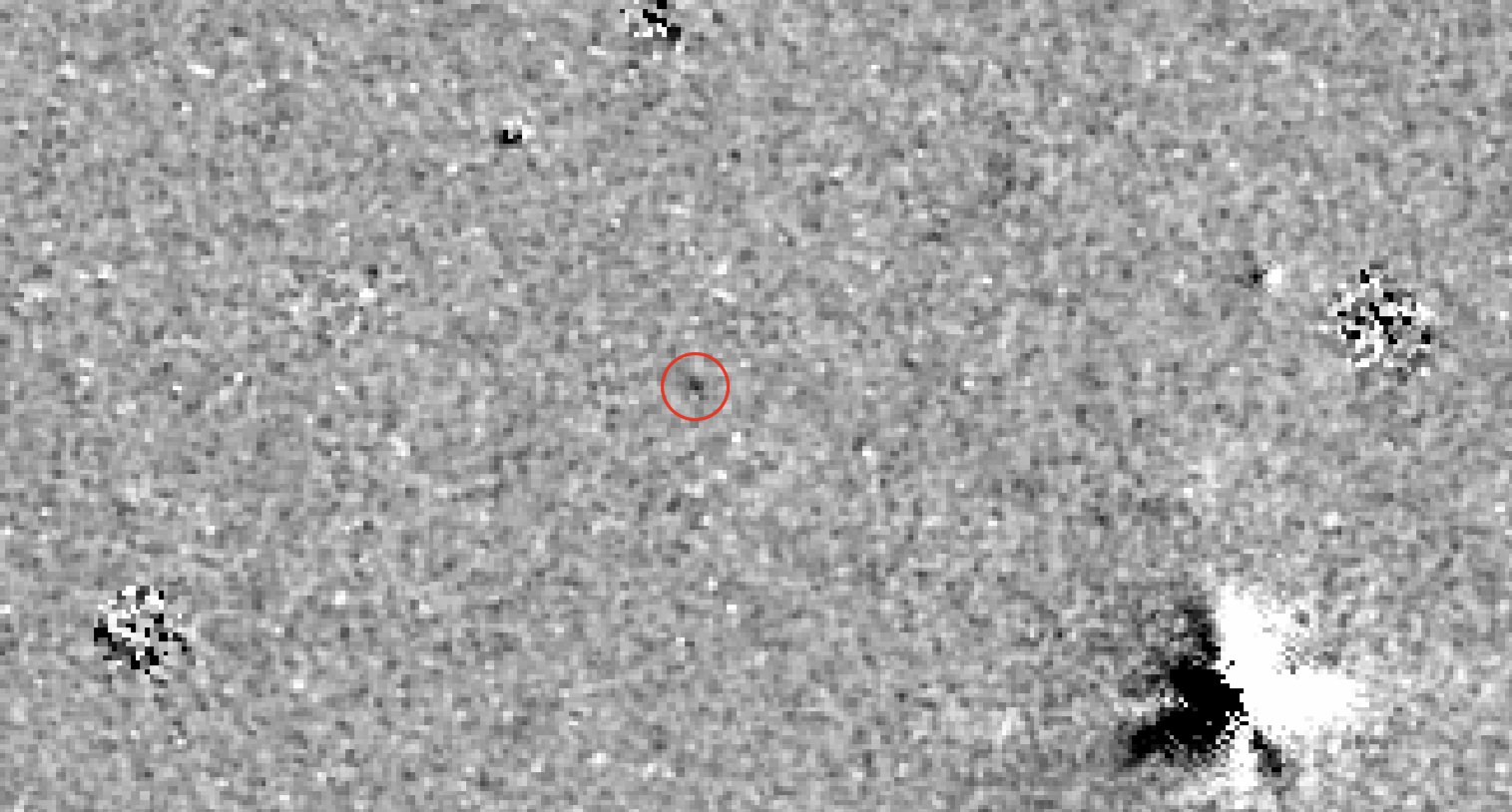}
        \includegraphics[width=0.33\linewidth]{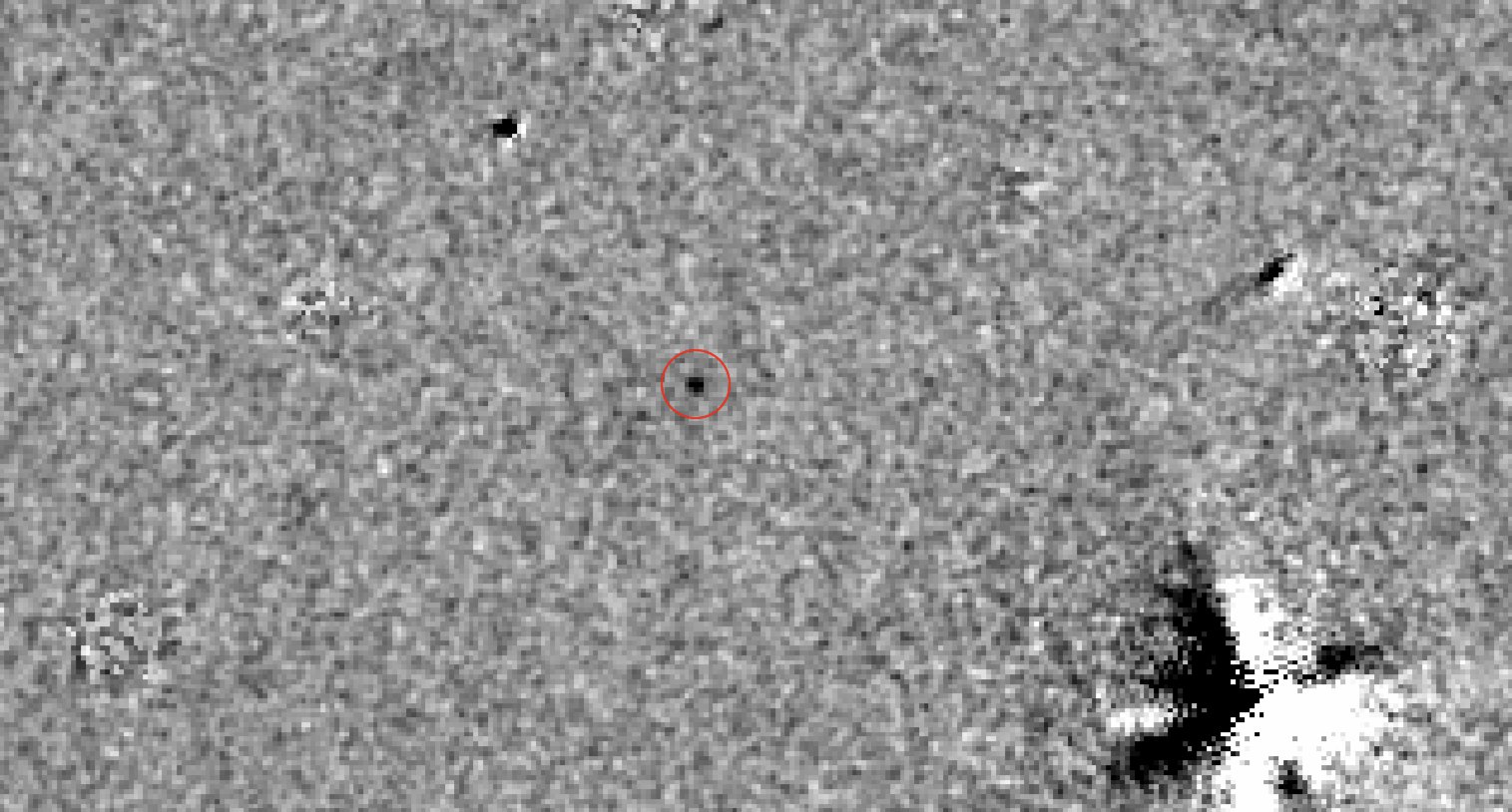}
    \end{subfigure}
    \caption{\emph{Top panel:} First and second epoch of our HST WFC3 F606W+F105W+F125W+F160W filters observations of the field of \fxt~combined. The red circle shows the location of the transient first identified in our NOT observations. In addition, a galaxy to the North of the lenticular galaxy is indicated with "N". This galaxy has a redshift of $z=2.18$ (see text). The figure also clearly shows the proximity of the optical counterpart of \fxt\,to the candidate (lenticular) host galaxy WISEA~J111002.65$-$075211.9 at $z=0.082$. Finally, an over-density of stars seemingly connecting the lenticular galaxy with the location of the transient is indicated with a green ellipse to guide the eye. \emph{Bottom left panel:} First epoch of our HST WFC3 F606W filter observations of the field of \fxt, showing the location of the transient first identified in our NOT observations at a magnitude of F606W=26.31$\pm$0.08, to be compared with the $r^\prime= 23.3$-band magnitude at detection.  
    \emph{Bottom middle panel:} The residual image resulting from subtracting the second epoch of our HST WFC3 F105W filter observation from the first. In the region indicated by the red circle a positive residual is present, indicating the counterpart faded between the first and second epoch of our HST F105W observations. 
    \emph{Bottom right panel:} The residual image resulting from subtracting the second epoch of our HST WFC3 F125W filter observation from the first. 
    The fading further solidifies the association between the counterpart and the FXT~\fxt. In all these three panels the size of the circle is the same (0.5\arcsec) as that in the \emph{top} panel. }
    \label{fig:hst}
\end{figure*}

The observations were taken over the periods 7.4--8.7~d and 28.8--29.0~d after the WXT trigger. Images were aligned to sources in common to each frame, and subsequently drizzled to final pixel scales of 0.05 and 0.07 arcseconds per pixel for the UVIS and IR filter observations, respectively. A source is detected in all HST filter observations in both epochs at the position of the NOT-optical counterpart. We combined all the filter observations and both epochs into a single image drizzled to a pixel scale of 0.15\arcsec~(see the \emph{top panel} of Fig.~\ref{fig:hst}). An over-density of stars seemingly connecting the lenticular galaxy with the location of the transient is found and indicated with a green ellipse to guide the eye in the figure. In addition, in Fig.~\ref{fig:hst} we show the resultant first epoch F606W image (\emph{Bottom left panel}), the 
difference image obtained subtracting the second epoch F105W image from the first epoch (\emph{Bottom middle panel}), and the difference image obtained subtracting the second epoch F125W image from the first epoch ({\emph{Bottom right panel}}). 

From the difference images it is clear that the source faded between the first and second epochs of HST observations in the F105W and F125W bands, while the magnitude measurements in the F606W bands also show it faded in F606W. Due to the larger measurement uncertainties in the F160W band observations (see Table~\ref{tab:photometry}) there is only marginal evidence for fading between HST epoch 1 and 2 in that band.  We collate all the optical and NIR photometry in Table~\ref{tab:photometry} and we show the light curves in Fig.~\ref{fig:lc}.

\subsubsection{Very Large Telescope/MUSE}
\label{sec_muse}
We observed the candidate host galaxy and the location of \fxt\, on 2025-03-03 using the Multi Unit Spectroscopic Explorer (MUSE) mounted on Very Large Telescope (VLT) Unit Telescope 4. The observations are part of the program with ID: 111.259Q (PI Jonker) and started at 02:47:36 (UTC) and lasted until 03:50:08 (UTC). Four exposures of 697~s each were obtained with small positional offsets between the exposures, however, the seeing deteriorated significantly while the exposures were obtained. Therefore, we only used the first two exposures with the best seeing of $\approx$1\arcsec. The data are reduced using the ESO Reflex pipeline \citep{Weilbacher2020,ESO2015}.

\begin{figure}
    \centering
    \includegraphics[width=0.95\linewidth]{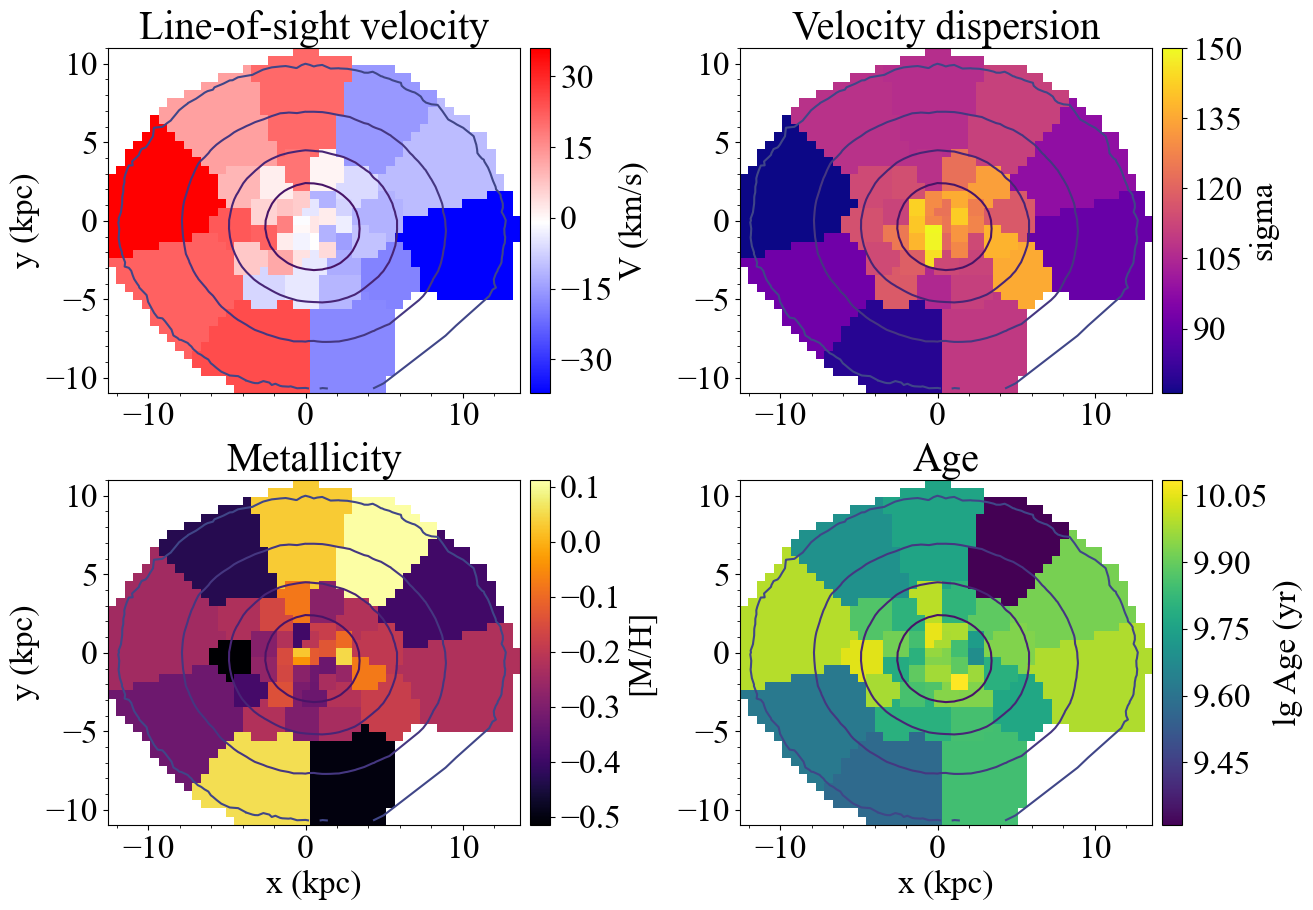}
    \caption{Spatial distributions of the the line-of-sight velocity \textit{V}, the velocity dispersion $\sigma$, the metallicity [M/H] and age in the 45 Voronoi bins of the candidate host galaxy WISEA~J111002.65$-$075211.9 observed with MUSE. The axes are given in \textit{x} and \textit{y} distances from the central pixel of the $z=0.082$ galaxy in kiloparsecs. Note that the location of the transients lies outside the sky region shown.}
    \label{fig:muse_ppxf_bins}
\end{figure}

The MUSE cube data of the candidate host galaxy is spatially Voronoi binned to a target signal-to-noise ratio (S/N)$=30$ per bin using the \texttt{VorBin} method and software of \citet{2003Cappellari}. Each spectroscopic bin in each spaxel with a $S/N<5$ is rejected to remove residual-dominated spectra before binning. Out of the 46 spatial spectra, one is contaminated by light of an object to the South-East of the galaxy, therefore this spatial bin is removed, leaving 45 (see Fig.~\ref{fig:muse_ppxf_bins}).  

The candidate host galaxy, WISEA J111002.65$-$075211.9 (\citealt{2025GCN.39278....1L}) is classified in NED as an irregular spiral galaxy, however in the HST images it visually resembles a lenticular or elliptical galaxy. In Fig.~\ref{fig:muse_ppxf_bins}, we show the spatial distributions for the radial velocity \textit{V}, the velocity dispersion $\sigma$, the total metallicity [M/H], and age of the stellar population. We note that the spatial variation detected in \textit{V}, shown in the top-left panel of Fig.~\ref{fig:muse_ppxf_bins} is typical for that observed in a lenticular galaxy (e.g.,~\citealt{2004MNRAS.352..721E}). 

\begin{figure*}
    \centering
    \includegraphics[width=0.95\linewidth]{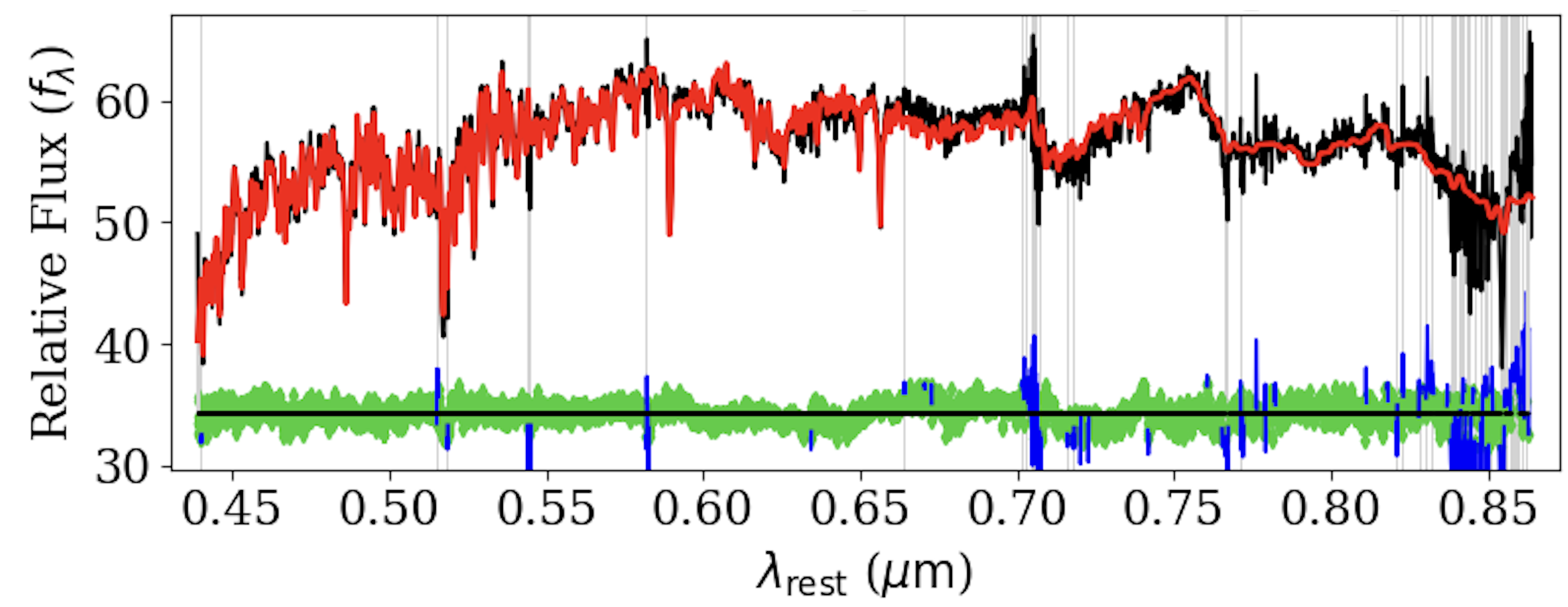}
    \caption{Best-fit \texttt{pPXF} (\citealt{2017Cappellari}) model to the MUSE spectrum of the host galaxy. The observed spectrum is shown in black, with the flexible stellar population synthesis template in red. Model-subtracted residuals are shown in green, with blue and the grey shaded regions showing the removed outliers to the model.}
    \label{fig:muse_ppxf}
\end{figure*}

We also obtain the average spectrum of the whole galaxy combining the 45 spatial bins. We use the penalized pixel fitting method (\texttt{pPXF}; \citealt{2017Cappellari}) to fit the spectrum. We use \textit{Flexible Stellar Population Synthesis} (fsps v3.2; \citealt{2009ConroyI, 2010ConroyII, 2010ConroyIII}) as the template bank for our stellar population synthesis model. As input parameters, we use the redshift of \textit{z}=0.082, $v=0$~km s$^{-1}$ (with respect to this redshift), and stellar velocity dispersion of $\sigma=200$~km s$^{-1}$ as initial guesses. The best fit for the average spectrum of the galaxy is displayed in  Fig.~\ref{fig:muse_ppxf}. The average spectrum is shown in black, with the best-fit galaxy template over-plotted in red. The residuals of the fit are shown in green.

\subsection{MeerKAT Radio observations}
We observed the position of \fxt~with the MeerKAT radio telescope \citep{Camilo2018, Jonas2018}, as part of program SCI-20241101-FC-01 (PI Carotenuto). We conducted three observations log-spaced in time, each with the same total on-source time of 42 minutes. The first observation started on February 13, 2025 at 01:02 UTC ($\approx$5.1 days after the first X-ray detection). The second and third observations were performed, respectively, on March 3, 2025 at 00:47 UTC ($\approx23$ days after the first X-ray detection) and on March 23, 2025 at 00:41 UTC ($\approx43$ days after the first X-ray detection). We observed at a central frequency of 3.06\,GHz (S-band, S4), with a total bandwidth of 875\,MHz. PKS~J1939--6342 and PKS~1128–047 were used as flux and complex gain calibrators, respectively. The data were reduced with the \texttt{OxKAT} pipeline \citep{oxkat}, which performs standard flagging, calibration and imaging using \texttt{tricolour} \citep{Hugo_2022}, \texttt{CASA} \citep{CASA_team_2022} and \texttt{WSCLEAN} \citep{Offringa_wsclean}, respectively. In the imaging step, we adopted a Briggs weighting scheme with a $-0.3$ robust parameter, yielding a $\sim 3.6\arcsec$ beam and a $8 \, \mu$Jy\,beam$^{-1}$ rms noise in the target field.  We do not detect radio emission at the position of the optical counterpart of EP250207b, and we place a 3$\sigma$ upper limits on the peak flux density of the target at 27, 24, and 27~$\mu$Jy beam$^{-1}$, for the first, second, and third observations, respectively. To search for persistent radio emission at the location of the source, we also stacked the data of the three epochs. No radio source was detected down to a 3$\sigma$ upper limit of 15~$\mu$Jy beam$^{-1}$.

\section{Discussion}
After the EP discovery of \fxt, we obtained X-ray, optical, NIR, and radio observations to follow the source evolution. We could not derive a direct redshift of the transient. However, the transient's location, next to the large lenticular galaxy WISEA~J111002.65$-$075211.9, provides statistical evidence through the low P$_{\textrm {chance}}<0.5$\% (\citealt{2025GCN.39270....1M}) for their association. 
Furthermore, our combined, deep, HST observations show that the source position is consistent with the outskirts of this galaxy (Fig.~\ref{fig:hst}). This image further reveals the presence of enhanced emission seemingly bridging the lenticular galaxy and the position of the transient (see the green ellipse in the top panel of Fig.~\ref{fig:hst}), further strengthening the suggested physical connection. Finally, in our deep HST imaging, we find no evidence for extended emission at the source position, such as could be the case had the source originated in a not-too-distant background galaxy. Therefore, we first discuss the nature and the properties of the source under the assumption that it lies at the same redshift as the lenticular galaxy ($z$=0.082; \citealt{2025GCN.39278....1L}; and this work, see Section~\ref{sec_muse}).

\begin{figure}
\centering
\includegraphics[width=0.99\columnwidth]{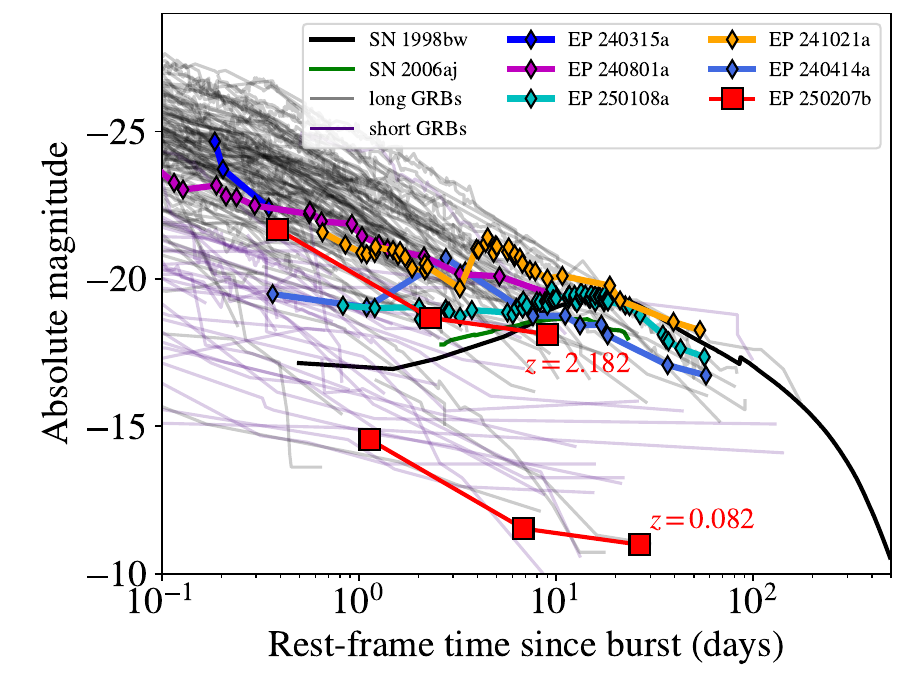}
\caption{Kann plot showing the approximate $r^\prime$-band light curve of \fxt\,in red together with those of a sample of EP-discovered FXTs (EP240414a [blue; \citealt{van_Dalen2024}], EP241021a [orange; Quirola-V\'asquez et al.~submitted], EP240801 [purple; \citealt{2025arXiv250304306J}], EP240315a [dark blue; \citealt{Levan2024}], 
EP250108a [magenta; \citealt{2025arXiv250408886E, 2025arXiv250408889R}]) and short GRBs (thin purple lines) and long GRBs (thin black lines). The GRB light curves are obtained from \citet{Kann2006,Kann2010,Kann2011, Nicuesa2012}. The absolute magnitudes of the data-points in the light curve from \fxt\, at $z=0.082$ are consistent with the faint-end of that of the short GRB distribution. If the source redshift is $z=2.18$ instead (see Discussion), the light curve is consistent with that of the bright end of short GRBs. The evolution as a function of time is consistent with that seen in short GRBs also, although the decay rate seems to decelerate after the first HST observation (near $t=7-8$~d). For a redshift of $z=0.082$, the presence of a (broad-lined) Type Ic (or indeed any) supernova can be ruled out. }
\label{fig:kann}
\end{figure}

At a redshift of $z$=0.082, the observed brightest absolute magnitude of $r_{\mathrm{ABS}}^\prime=-14.5$ is rather faint when compared to the peak absolute magnitude of several EP-discovered FXTs (e.g.,~see Fig.~\ref{fig:kann}).  However, while it is too faint to be explained as an afterglow of a long GRB, it is consistent with the peak absolute magnitude and light curve evolution of some fainter, merger-driven short GRBs (Fig.~\ref{fig:kann}). Furthermore, the offset of 10\arcsec~(\citealt{2025GCN.39300....1L}), corresponding to 15.9~kpc in projection, is well within the range of typical host galaxy offsets observed for short GRBs (e.g., \citealt{2002AJ....123.1111B,2022ApJ...940...56F}) and simulated merger origin GRB population studies \citep{mandhai2022}. In addition, the high age of the stars in the lenticular galaxy that we derive from our VLT/MUSE observations (see Fig.~\ref{fig:muse_ppxf_bins}) is inconsistent with a collapsar origin, but is consistent with a merger-driven (short) GRB scenario. Finally, the rate of decay observed in the $r^\prime$ and F606W-band (see Fig.~\ref{fig:lc}) seems to decelerate. This could be consistent with persistent contributions from a globular cluster or the core of a (tidally disrupted) dwarf galaxy host for \fxt. At a redshift of $z=0.082$, a globular cluster near the peak of the absolute magnitude distribution, i.e., with an absolute magnitude of $\approx -10$ (\citealt{Harris2010}), would correspond to F606W=27.84 AB mag and thus would contribute about 30\% to the observed flux at our observational epoch $t\approx 28$~d. Clearly, an even brighter globular cluster could be responsible for nearly all the optical/NIR light in this final epoch. The absolute magnitude distribution of ultra compact dwarfs (UCDs), which might be the cores of a tidally disrupted dwarf galaxies, overlaps that of the bright end of the absolute magnitude distribution of globular clusters (e.g.~\citealt{mieske2002}), so the source observed at late time could also be explained by such a UCD.  A tidal stream from a tidally disrupted dwarf galaxy could also explain the enhanced emission linking WISEA~J111002.65$-$075211.9 and the location of the transient. 
Future HST or JWST observations of \fxt\ can test the scenario that part or all of the light in the last epoch of HST observations is due to a globular cluster or a disrupted dwarf galaxy. Finally, for a redshift of $z=0.082$, the absolute magnitudes on rest frame timescales of $\approx 5$ to $25$~d after the discovery of \fxt~rule out the presence of a (broad-lined) Type Ic (or indeed any) supernova.

\begin{figure}
\centering
\includegraphics[width=0.99\columnwidth]{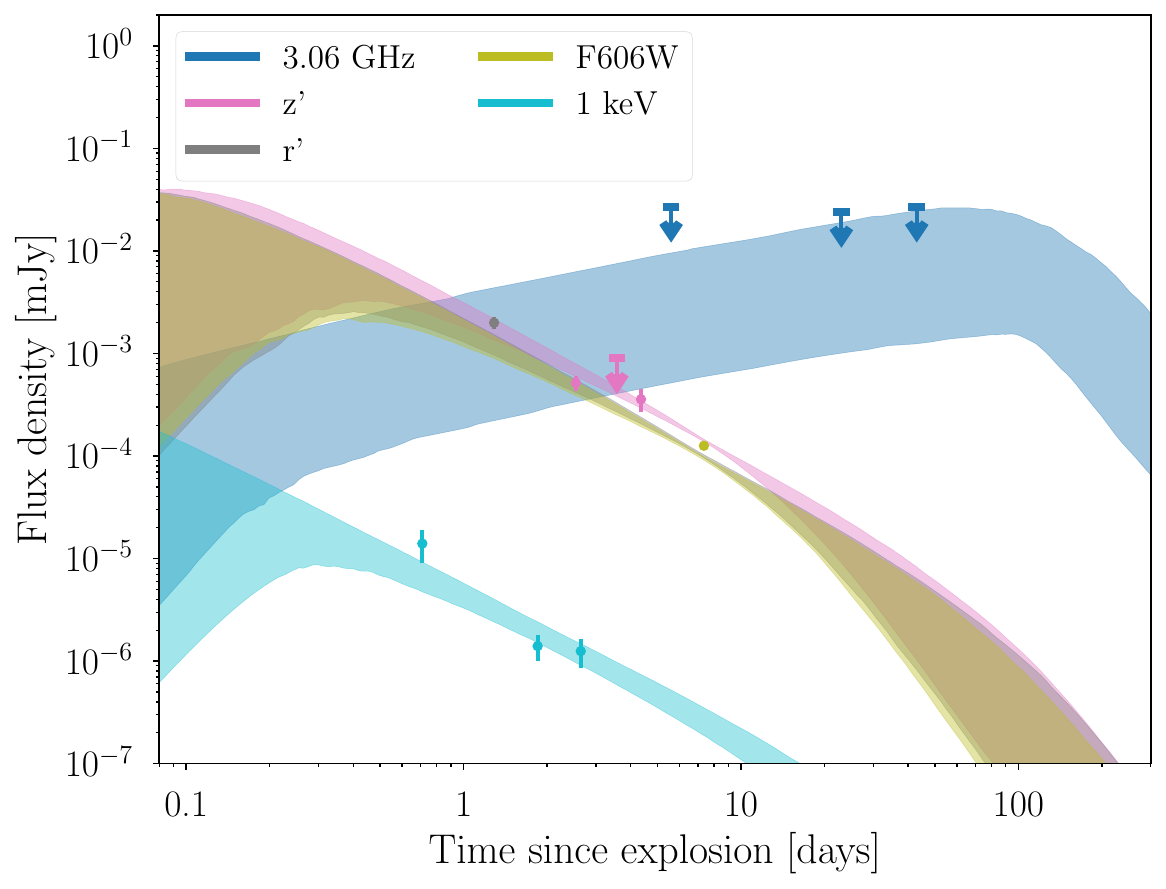}
\caption{We used {\sc redback} (\citealt{Nikhilredback2024}) to fit the optical (grey, green and pink) and X-ray (light blue) afterglow data (excluding the data obtained after $t>10$~d), including the radio (blue) upper-limits at an assumed redshift, $z = 0.082$. The afterglow model used was {\texttt gaussian\_redback} following \citet[][]{lamb2018} and including synchrotron self-absorption effects \citep[see][for details]{lamb2019c}. The shaded regions indicate the 90\% credible interval for the model fits to the data, where we used {\sc nessai} \citep{wiiliams2024} as the sampler with a Gaussian likelihood. The model fits return an out-of-jet-core viewing angle, essentially perhaps slightly `off-axis', of $\sim 6\mathring{.}3$ for a jet core angle of $\sim4\mathring{.}6$, however, the uncertainties on both parameters are also consistent with an 'on-axis' scenario. Furthermore, the Gaussian structured jet `wings' extend to $\sim9\mathring{.}2$, although the energy at the wider angles will contribute insignificantly to the observed emission. After the initial decay from the prompt emission (the prompt emission is not modelled here) the X-ray observations from EP-FXT, the optical/NIR data at $t<10$~d, and radio upper limits are consistent with mildly off-axis afterglow emission.}
\label{fig:afpy}
\end{figure}

The 3.06~GHz radio MeerKAT non-detections at $t=$5.6, 23, and 43~d after the discovery at flux levels of $\approx25\mu$Jy imply a radio luminosity upper limit of $L_{3.06~\mathrm{GHz}}\approxlt 1.3\times 10^{37}$~\lum, or $\approxlt 4\times 10^{27}$~\lum\,Hz$^{-1}$. Such a radio luminosity is low when compared to the radio emission detected for (on-axis) short GRBs (see e.g.~figure 13 in \citealt{2021ApJ...906..127F}).
We used {\sc redback} (\citealt{Nikhilredback2024}) to compare our radio, optical ($r^\prime$/V-band), and X-ray light curves with those estimated via the {\texttt afterglow\_models.gaussian\_redback} structured jet model. 
This model is identical to that used for the only confirmed `off-axis' viewed merger origin GRB\,170817A \citep[e.g.,][]{lyman2018,lamb2019b}, and  assumes a Gaussian-shaped jet structure.
The external medium of the afterglow is assumed to be uniform, and the jet undergoes sideways expansion as described by \cite{granot2012} for their $a = 1$ case.
The model priors and posterior distributions are presented in Table\,\ref{tab:prior-posterior} and the fit is indicative of a mildly `off-axis' viewed GRB afterglow in a low-density ambient medium, see Figure\,\ref{fig:afpy}. Note that \citet{Wichern2024} showed that an off-axis scenario is difficult to reconcile with the properties of the full sample of  \chan-discovered FXTs, suggesting perhaps that (some) EP-discovered FXT have a different nature than those. Finally, the optical emission detected in the second HST epoch of observations is too bright for our afterglow model, suggesting that additional emission mechanisms or sources (such as the possible globular cluster or dwarf galaxy mentioned above) could contribute to the optical light.

\begin{table*}
    \centering
    \begin{tabular}{c|c|c|c}
       Parameter  & Prior  & Posterior & Description\\
       \hline
        & & & \\
       $\theta_{\rm observer}$ (rad) & $[\sin] 0 \leftrightarrow \pi/2$ & $0.11\pm0.04$ & Observers line-of-sight angle \\
       $\log E_{\rm K, iso}$ ($\log$ erg) & $44 \leftrightarrow 54$ & $51.7\pm0.7$  & Isotropic equivalent kinetic energy \\
       $\theta_{\rm core}$ (rad) & $0.01 \leftrightarrow 0.1$ & $0.08\pm0.02$  & Jet core half opening angle \\
       $\theta_{\rm edge}$ (rad) & $0.1 \leftrightarrow 0.2$ & $0.16\pm0.03$ & Angular extent of structured jet \\
       $\log n_{\rm ism}$ ($\log$ cm$^{-3}$) & $-5 \leftrightarrow 2$ & $-4.3^{+0.8}_{-0.5}$  & Ambient medium number density \\
       $p$ & $1.4\leftrightarrow3.1$ & $2.92\pm0.06$ & Electron spectral energy density index \\
       $\log\epsilon_e$ & $-5\leftrightarrow0^\dagger$ & $-0.4^{+0.2}_{-0.4}$ & Fraction of energy in electrons \\
       $\log\epsilon_B$ & $-5\leftrightarrow0^\dagger$ & $-2.9\pm1.4$ & Fraction of energy in magnetic field \\
       $\varepsilon_N$ & $0.1\leftrightarrow1.0$ & $0.34^{+0.3}_{-0.17}$ & Synchrotron participation fraction \\
       $\Gamma_0$ & $40 \leftrightarrow400$ & $260\pm110$ & Initial bulk Lorentz factor \\
       & & & 
    \end{tabular}
    \caption{The model parameters for our afterglow model fit data shown in Figure\,\ref{fig:afpy}. Note that we excluded the optical and NIR data obtained after $t>10$~d from the fit, as this emission seems to be coming from a component that is not an afterglow. The prior range and, where appropriate, the distribution function are given (else the distribution is flat in the range indicated), and the model posterior median and $1-\sigma$ confidence interval. A corner plot of the posterior distribution is included in Appendix\,\ref{ap:post}. $^\dagger$We also set the requirement $\log\epsilon_e>\log\epsilon_B$.}
    \label{tab:prior-posterior}
\end{table*}

The average 0.3--10 keV X-ray luminosity at the EP-WXT discovery is $L_{{\textrm X, ave}}=1\times 10^{46}$~\lum. This X-ray luminosity is on the high end, but still consistent with, the X-ray luminosity theorized to be emitted through the spin-down of a ms massive ($>2$~M$_\odot$) magnetar formed in a binary neutron star merger (e.g.~\citealt{2013ApJ...763L..22Z,2014MNRAS.439.3916M,2016ApJ...829...72C,Sun2017, Sun2019,Quirola2024}). Note that under this model the X-ray emission is quasi-isotropic and not powered by a jet. However, the X-ray light curve can be well described by a simple power-law decay, i.e., no clear, ks-lasting plateau in the light curve is detected, such as has been previously invoked for  FXTs that are suggested to be magnetar-powered (e.g.~\citealt{2013ApJ...763L..22Z}; \citealt{2014MNRAS.439.3916M}; \citealt{2016ApJ...829...72C}; \citealt{2017ApJ...835....7S}; \citealt{2019ApJ...886..129S}; \citealt{Xue2019}; \citealt{Quirola2024}). Although, as the EP-WXT observation was cut short (the transient was still ongoing when the satellite started moving), the measured FXT duration is a lower limit. Nevertheless, it is possible that our viewing angle to \fxt\ is such that most of the plateau emission is not observed (the so-called "trapped zone"; \citealt[e.g.,][]{2019ApJ...886..129S}). Only after the X-ray emission has ionised the ejecta in our line of sight the X-rays escape, leading to the detection of, in this case, a small part of the plateau (describing the EP-WXT light curve with a constant; see Fig.~\ref{fig:lc-wxt}) and the power law decay phase (cf.~\citealt{2019ApJ...886..129S}; see Fig.~\ref{fig:lc}). 

The index of the best-fit power law decay of the X-ray light curve of $-1.5$ is in-between the predictions for the decay of $L_X\propto t^{-1}$ and $L_X\propto t^{-2}$ for times larger than the characteristic timescale for magnetar spin down due to the emission of gravitational wave radiation and for times larger than the characteristic timescale for magnetar spin down caused by electromagnetic radiation, respectively (cf.~\citealt{Quirola2024}). However, the power-law index does not need to be $-1$ or $-2$, if the electromagnetic radiation from a ms magnetar is not only from vacuum dipole radiation. Moreover, the spin-down mechanism evolves with time. Therefore, given that the value of $\approx -1.5$ is measured over a days- to week-long time period, it can also reflect this evolution (see e.g., \citealt{Lasky2017}, \citealt{Sarin2020}).

\begin{figure*}
\centering
\includegraphics[width=2.\columnwidth]{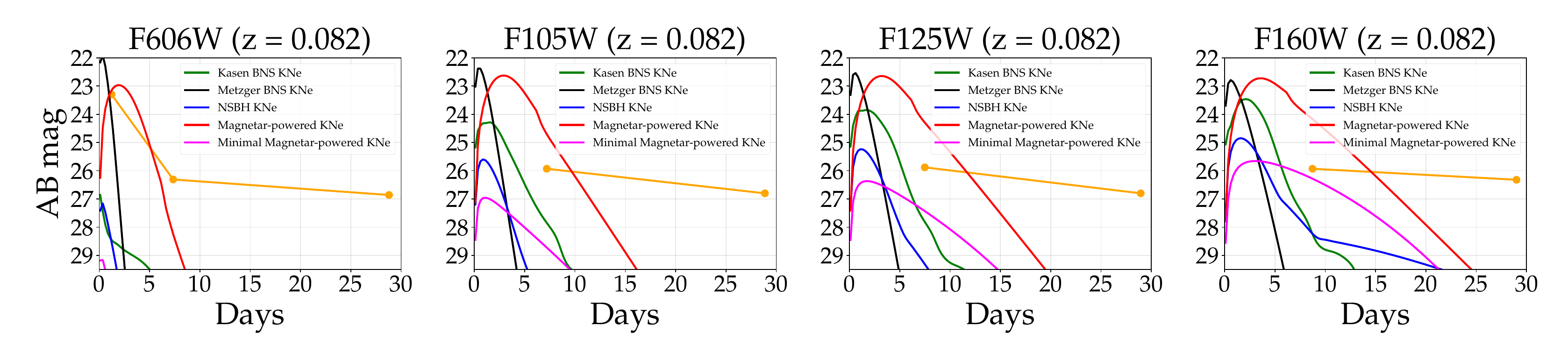}
\caption{Plotted are the optical NIR data point in the filters as indicated and light curves for different KN models in the four main filters used in this work assuming a of redshift $z=0.082$. We follow \citet{Kasen2017} and  \citet{Metzger2017LRR} for KN models, and we follow \citet{Nikhil2022magne} for a magnetar enhanced KN  and a black hole -- neutron star merger model. We use these models as implemented in {\sc Redback} (\citealt{Nikhilredback2024}; {\sc two\_component\_nsbh\_ejecta\_relation} for the  black hole -- neutron star merger model) to calculate theoretical KN light curves in the  optical ($r^\prime$/F606W) and NIR (F105W, F125W, F160W) filters. We find that if \fxt\, is indeed at $z=0.082$, only KN emission models with relatively low ejecta masses are consistent with the observed magnitudes (yellow symbols) at $t\approx1.23, 7.4-8.7$~d. A typical magnetar-enhanced KN model is over-predicting the KN emission compared with the HST/WFC3 NIR data (red curve). Only an extreme version of a magnetar enhanced KN is consistent with the data (magenta curve). The two data points at $\approx7/28$~d in each panel are the two HST measurements, the first data point in the \emph{Left} panel is the NOT $r^\prime$-band measurement. Note that while the source is too bright to be explained by kilonova emission during the second epoch ($t\approx 28-29$~d), this could be explained if the source resides in a globular cluster or the core of a (tidally disrupted) dwarf galaxy. See text for details.}
\label{fig:kn}
\end{figure*}

We have used several kilonova (KN) models implemented in {\sc Redback} (\citealt{Nikhilredback2024}) to calculate possible KN optical (F606W) and NIR (F105W, F125W, F160W) light curves to compare with the HST data points (see Fig.\ref{fig:kn}). We find that many of the KN models from \citet{Kasen2017}  over-predict the observed HST magnitudes at observer times 7.3--8.7~d after the source discovery. Only KN emission produced by mergers giving rise to a relatively low ejecta mass of 0.005~M$_\odot$ is consistent with the observations (green line in Fig.~\ref{fig:kn}). The assumed lanthanide fraction is 10$^{-1}$. Such ejecta masses are low compared to the ejecta masses typically found from modelling the observations of GW~170817 (e.g.~\citealt{Villar2017}, \citealt{2018MNRAS.481.3423W}) and samples of short GRBs (e.g.~\citealt{Rastinejad2021}). Also, while there are not many spectroscopically confirmed KNe known, such low ejecta masses are rarely seen in numerical simulations which produce hypermassive or longer-lived neutron stars (\citealt{Kawaguchi2022, Kawana2018}), i.e., the magnetars that have been invoked to explain some FXTs, and more likely points towards a binary neutron star merger where the remnant promptly collapsed into a black hole~\citep{Nedora2022}. The latter tend to produce less ejecta, both dynamically and from the disk-winds~(e.g., \citealt{Siegel2019, Sarin2021review}). 

The KN models from \citet{Metzger2017LRR} produce a bright KN signal that rapidly decays, and can just be consistent with the observations; e.g., for an ejecta mass of 0.01~M$_\odot$, ejecta velocity of 0.25~$c$, lanthanide fraction $\chi$=0.1, velocity index $\beta=3$ for $m_{{\textrm ejecta}}\propto v^{-\beta}$, and a grey-opacity of $\kappa=0.5$ corresponding to lanthanide-poor, ``blue'' ejecta. Magnetar enhanced KN models from \citet{Nikhil2022magne} for typical parameter values over-predict the optical and NIR magnitudes (red line in Fig.~\ref{fig:kn}). The magnetar enhanced models can only be made consistent with the HST observations for rather constraining, probably even implausible, parameters (magenta line in Fig.~\ref{fig:kn}). This model requires significant gravitational-wave emission, which reduces the available rotational energy budget available for electromagnetic radiation, has a high $\gamma$-ray opacity to further reduce the available energy in optical, and finally, we force the opacity to be high, in contrast to expected opacities in the case of long-lived neutron star remnants~(\citealt{Metzger2019}). 

Finally for a redshift of $z=0.082$, we calculate using {\sc bilby} (\citealt{2019ApJS..241...27A}) that any binary neutron star merger like GW~170817 would probably provide a marginal gravitational wave (GW) signal detection with a signal-to-noise ratio (SNR) of $\approx9$ in LIGO-Virgo Kagra (LVK), i.e., below the conventional trigger threshold of an SNR 12. If the event was instead due to a black hole -- neutron star merger, the black hole mass and spin would have to have been such that there was (ample) of material outside the black hole's event horizon, implying that as the GW signal would be stronger than for a BNS merger, it might well have yielded a detection. The lack of such a detection suggests therefore that a black hole -- neutron star merger is less likely as an origin for \fxt. Combining the time and sky location of \fxt~with GW data will boost the confidence in any GW signal (cf.~\citealt{2023MNRAS.518.5483S}).

Next, we consider an alternative scenario where the association with the candidate host galaxy, WISEA~J111002.65$-$075211.9 is a mere chance alignment and the FXT must originate from (near) another host galaxy. The light detected in the second-epoch of HST observations could be due to an unresolved background galaxy. At a magnitude of F606W~$\approx27$ and an offset of $\approxlt 1$ arcsecond, the chance alignment probability for such a galaxy is $\approx10$\% (\citealt{bloom2002}), significantly larger than the chance alignment probability of $\approxlt0.5$~\% for WISEA~J111002.65$-$075211.9 (\citealt{2025GCN.39270....1M}). Nevertheless, if we assume the second epoch HST detections of the source are in fact due to the unresolved host galaxy, and we fit BAGPIPES (\citealt{bagpipes}) and Prospector (\citealt{prospector2021})  models to it we obtain a photometric redshift of $z_{\mathrm{BP}}=3.0\pm0.5$, and $z_{\mathrm{Prosp}}=3.5\pm0.7$. We have added a figure showing the best-fit Prospector galaxy model to the Appendix (\ref{ap:prosp}). For a redshift of $z=3$, the EP-WXT 0.3--10 keV X-ray luminosity would become $2\times10^{50}$~\lum, which is a typical value for long GRBs (e.g., see \citealt{2016ApJ...825L..20D}). The absolute magnitude would become restframe $g^\prime$-band$~\approx-21.5$, not out of the ordinary for a long GRB host galaxy at such a redshift (e.g., \citealt{2022A&A...666A..14S}). Thus, besides the higher chance alignment, this scenario cannot be ruled out. 

Finally, we also briefly consider as host of \fxt~the galaxy marked with a "N" in Fig.~\ref{fig:hst} that lies in projection to the North of the candidate host lenticular galaxy. Using the Legacy Survey (\citealt{dey2019}) magnitude converted to Vega mag and correcting for Galactic extinction we have R = 23.7 at a separation of 5.8” with respect to the FXT. This leads to chance association probability of 32\% for the galaxy. Even though this is thus a likely chance alignment, we derived the redshift from its MUSE spectrum. Two clear emission lines are found at observed wavelengths of 8132.2$\pm0.3$ and 8138.6$\pm$0.4\AA. The wavelengths of these lines are consistent with those of the [OII] doublet at a redshift of $z=2.1824$. If \fxt\, and its counterpart are in fact associated with this galaxy, the offset of the optical counterpart to the centre of this galaxy of $\approx$5.78\arcsec\ corresponds to an offset of about 50~kpc in projection, still consistent with the distribution of offsets found for short GRBs (see references above) but inconsistent with the cumulative distribution for long GRBs (\citealt{2016ApJ...817..144B}). The observed peak absolute magnitude in the $r^\prime$-band would become $-21.7$, near the bright-end of the short GRB distribution (see Fig.~\ref{fig:kann}) and the average X-ray luminosity at the EP-WXT discovery $L_{{\textrm X}}=2\times 10^{49}$~\lum\, for this higher redshift. The radio $L_{3.06~\textrm{GHz}}$ would become $\approxlt 3\times 10^{40}$~\lum, or $\approxlt 9\times 10^{30}$~\lum\,Hz$^{-1}$. These values for the radio luminosity are in line with those found for other short GRBs (see e.g.~figure 13 in \citealt{2021ApJ...906..127F}). However, the peak X-ray luminosity in this case becomes too high to be a magnetar-powered FXT and we must be observing the jet-related prompt X-ray emission. Concluding, even if the galaxy marked "N" is the host galaxy, this EP-discovered FXT can be explained as due to a merger driven event and not as a collapsar-driven event.

\section*{Acknowledgements}
We thank the referee for their comments which helped improve the Manuscript.

We are deeply grateful to Tom Marsh for developing the molly software, one of his many contributions to advancing the field of compact objects. 

P.G.J. J.N.D.D., J.S.S., J.Q.V., M.E.R. and A.P.C.H.~are supported  by the European Union (ERC, Starstruck, 101095973, PI Jonker). Views and opinions expressed are however those of the author(s) only and do not necessarily reflect those of the European Union or the European Research Council Executive Agency. Neither the European Union nor the granting authority can be held responsible for them.
D.M.S. and M.A.P.T. acknowledge support by the Spanish Ministry of Science via the Plan de Generacion de conocimiento PID2020-120323GB-I00. DMS also acknowledges support via a Ramon y Cajal Fellowship RYC2023-044941. J.Q.V. additionally acknowledges support by the IAU-Gruber foundation. SJS acknowledges funding from STFC Grant ST/Y001605/1, a Royal Society Research Professorship and the Hintze Charitable Foundation. D.B.M.~is funded by the European Union (ERC, HEAVYMETAL, 101071865). The Cosmic Dawn Center (DAWN) is funded by the Danish National Research Foundation under grant DNRF140.
T.-W.C. acknowledges the financial support from the Yushan Fellow Program by the Ministry of Education, Taiwan (MOE-111-YSFMS-0008-001-P1) and the National Science and Technology Council, Taiwan (NSTC grant 114-2112-M-008-021-MY3).
F.E.B. acknowledge support from ANID-Chile BASAL CATA FB210003, FONDECYT Regular 1241005,
and Millennium Science Initiative, AIM23-0001.
GL acknowledges support from a VILLUM FONDEN research grant (VIL60862). 
GPL is supported by a Royal Society Dorothy Hodgkin Fellowship (grant Nos. DHF-R1-221175 and DHF-ERE-221005).
BPG acknowledges support from STFC grant No. ST/Y002253/1 and from The Leverhulme Trust grant No. RPG-2024-117. 
FO acknowledges support from MIUR, PRIN 2020 (grant 2020KB33TP) ``Multimessenger astronomy in the Einstein Telescope Era (METE)'' and from INAF-MINIGRANT (2023): "SeaTiDE - Searching for Tidal Disruption Events with ZTF: the Tidal Disruption Event population in the era of wide field surveys".

Based on observations obtained at the international Gemini Observatory (Program IDs GS-2024B-Q-131 and GS-2024B-FT-112), a program of NOIRLab, which is managed by the Association of Universities for Research in Astronomy (AURA) under a cooperative agreement with the National Science Foundation on behalf of the Gemini Observatory partnership: the National Science Foundation (United States), National Research Council (Canada), Agencia Nacional de Investigaci\'{o}n y Desarrollo (Chile), Ministerio de Ciencia, Tecnolog\'{i}a e Innovaci\'{o}n (Argentina), Minist\'{e}rio da Ci\^{e}ncia, Tecnologia, Inova\c{c}\~{o}es e Comunica\c{c}\~{o}es (Brazil), and Korea Astronomy and Space Science Institute (Republic of Korea). Data was processed using the Gemini DRAGONS (Data Reduction for Astronomy from Gemini Observatory North and South) package.

Data for this paper has in part been obtained under the International Time Programme of the CCI (International Scientific Committee of the Observatorios de Canarias
of the IAC) with the NOT and GTC operated on the island of La Palma by the Roque de los Muchachos.

The Legacy Surveys consist of three individual and complementary projects: the Dark Energy Camera Legacy Survey (DECaLS; Proposal ID \#2014B-0404; PIs: David Schlegel and Arjun Dey), the Beijing-Arizona Sky Survey (BASS; NOAO Prop. ID \#2015A-0801; PIs: Zhou Xu and Xiaohui Fan), and the Mayall $z$-band Legacy Survey (MzLS; Prop. ID \#2016A-0453; PI: Arjun Dey). DECaLS, BASS and MzLS together include data obtained, respectively, at the Blanco telescope, Cerro Tololo Inter-American Observatory, NSF’s NOIRLab; the Bok telescope, Steward Observatory, University of Arizona; and the Mayall telescope, Kitt Peak National Observatory, NOIRLab. Pipeline processing and analyses of the data were supported by NOIRLab and the Lawrence Berkeley National Laboratory (LBNL). The Legacy Surveys project is honored to be permitted to conduct astronomical research on Iolkam Du\'ag (Kitt Peak), a mountain with particular significance to the Tohono O\'odham Nation.

NOIRLab is operated by the Association of Universities for Research in Astronomy (AURA) under a cooperative agreement with the National Science Foundation. LBNL is managed by the Regents of the University of California under contract to the U.S. Department of Energy.

The data presented here were obtained [in part] with ALFOSC, which is provided by the Instituto de Astrofisica de Andalucia (IAA) under a joint agreement with the University of Copenhagen and NOT.

This project used data obtained with the Dark Energy Camera (DECam), which was constructed by the Dark Energy Survey (DES) collaboration. Funding for the DES Projects has been provided by the U.S. Department of Energy, the U.S. National Science Foundation, the Ministry of Science and Education of Spain, the Science and Technology Facilities Council of the United Kingdom, the Higher Education Funding Council for England, the National Center for Supercomputing Applications at the University of Illinois at Urbana-Champaign, the Kavli Institute of Cosmological Physics at the University of Chicago, Center for Cosmology and Astro-Particle Physics at the Ohio State University, the Mitchell Institute for Fundamental Physics and Astronomy at Texas A\&M University, Financiadora de Estudos e Projetos, Fundacao Carlos Chagas Filho de Amparo, Financiadora de Estudos e Projetos, Fundacao Carlos Chagas Filho de Amparo a Pesquisa do Estado do Rio de Janeiro, Conselho Nacional de Desenvolvimento Cientifico e Tecnologico and the Ministerio da Ciencia, Tecnologia e Inovacao, the Deutsche Forschungsgemeinschaft and the Collaborating Institutions in the Dark Energy Survey. The Collaborating Institutions are Argonne National Laboratory, the University of California at Santa Cruz, the University of Cambridge, Centro de Investigaciones Energeticas, Medioambientales y Tecnologicas-Madrid, the University of Chicago, University College London, the DES-Brazil Consortium, the University of Edinburgh, the Eidgenossische Technische Hochschule (ETH) Zurich, Fermi National Accelerator Laboratory, the University of Illinois at Urbana-Champaign, the Institut de Ciencies de l’Espai (IEEC/CSIC), the Institut de Fisica d’Altes Energies, Lawrence Berkeley National Laboratory, the Ludwig Maximilians Universitat Munchen and the associated Excellence Cluster Universe, the University of Michigan, NSF’s NOIRLab, the University of Nottingham, the Ohio State University, the University of Pennsylvania, the University of Portsmouth, SLAC National Accelerator Laboratory, Stanford University, the University of Sussex, and Texas A\&M University.

BASS is a key project of the Telescope Access Program (TAP), which has been funded by the National Astronomical Observatories of China, the Chinese Academy of Sciences (the Strategic Priority Research Program “The Emergence of Cosmological Structures” Grant \#~XDB09000000), and the Special Fund for Astronomy from the Ministry of Finance. The BASS is also supported by the External Cooperation Program of Chinese Academy of Sciences (Grant \#~114A11KYSB20160057), and Chinese National Natural Science Foundation (Grant \#~12120101003, \#~11433005).

The Legacy Survey team makes use of data products from the Near-Earth Object Wide-field Infrared Survey Explorer (NEOWISE), which is a project of the Jet Propulsion Laboratory/California Institute of Technology. NEOWISE is funded by the National Aeronautics and Space Administration.

The Legacy Surveys imaging of the DESI footprint is supported by the Director, Office of Science, Office of High Energy Physics of the U.S. Department of Energy under Contract No. DE-AC02-05CH1123, by the National Energy Research Scientific Computing Center, a DOE Office of Science User Facility under the same contract; and by the U.S. National Science Foundation, Division of Astronomical Sciences under Contract No. AST-0950945 to NOAO.

The MeerKAT telescope is operated by the South African Radio Astronomy Observatory, which is a facility of the National Research Foundation, an agency of the Department of Science and Innovation. This work has made use of the “MPIfR S-band receiver system” designed, constructed and maintained by funding of the MPI f\"ur Radioastronomy and the Max Planck Society.

\section*{Data Availability}

Data used in this paper is publicly available in the Zenodo repository 10.5281/zenodo.17579065 .

%
\bibliographystyle{mnras} 

\appendix

\section{Optical and NIR photometry}
\begin{table*}
\centering
\begin{tabular}{lcccccc}
\hline\hline
Telescope & Instrument & Date (UTC) & Days since trigger & Exposure time (s) & Filter & AB magnitude  \\ 
(1) & (2) & (3) & (4) & (5) & (6) & (7) \\ \hline
NOT & ALFOSC & 2025-02-09 03:16:13 & 1.228   & 4$\times$200 & $r^\prime$ & 23.3$\pm$0.16  \\
NOT & NOTCam & 2025-02-11 03:07:13 & 3.22 & 30$\times$60 & $J$ & $>22.8$  \\
GN & GMOS & 2025-02-10 10:45:59 & 2.54 & 6$\times$60 & $z^\prime$ & 24.7$\pm$0.2 \\
GN & GMOS & 2025-02-11 11:25:46  & 3.57 & 5$\times$60 & $g^\prime$ & $>$24.7  \\
GN & GMOS & 2025-02-11 11:27:33  & 3.57 & 5$\times$60 & $z^\prime$ & $>$24.1  \\
GS & GMOS & 2025-02-12 06:23:32 & 4.36 & 12$\times$60 & $z^\prime$ & 25.1$\pm$0.3  \\
GS & F2 & 2025-02-13 04:46:44 & 5.29 & 27$\times$40 & $J$ & $>$24.2 \\
GS & F2 & 2025-02-14 04:39:23  & 6.29 & 90$\times$15 & $K_s$ & $>$23.15  \\
HST & WFC3 & 2025-02-15 06:15:17  & 7.35 & 4$\times$505 & F606W & 26.31$\pm$0.08  \\
HST & WFC3 &  2025-02-15 07:50:29 & 7.42 & 4$\times$553 & F105W & 25.93$\pm$0.02  \\
HST & WFC3 & 2025-02-15 09:24:56  & 7.48 & 4$\times$553 & F125W & 25.88$\pm$0.11  \\
HST & WFC3 & 2025-02-16 15:19:42  & 8.73 & 4$\times$553 & F160W & 25.93$\pm$0.15  \\
HST & WFC3 & 2025-03-08 17:06:56  & 28.8 & 2$\times$505 & F606W & 26.86$\pm$0.15  \\
HST & WFC3 & 2025-03-08 18:41:49  & 28.87 & 4$\times$553 & F105W & 26.8$\pm$0.2  \\
HST & WFC3 & 2025-03-08 20:16:15  & 28.94 & 4$\times$553 & F125W & 26.8$\pm$0.2  \\
HST & WFC3 & 2025-03-08 21:50:40  & 29 & 4$\times$553 & F160W & 26.32$\pm$0.18  \\
\hline
\end{tabular}
\caption{Optical and near-infrared (NIR) photometry obtained through our follow-up observations. An $>$ in front of the number in the AB magnitude column indicates a 3$\sigma$ upper limit. The reported photometry is uncorrected for extinction, which taking the $\mathrm{N_H}=4\times 10^{20}$~cm$^{-2}$ from the X-ray spectral fits, would correspond to an $\mathrm{A_V}=0.18$~mag following \citet{2009MNRAS.400.2050G}.}
\label{tab:photometry}
\end{table*}

\section{Afterglow modelling posterior distribution}\label{ap:post}
\begin{figure*}
    \centering
    \includegraphics[width=\textwidth]{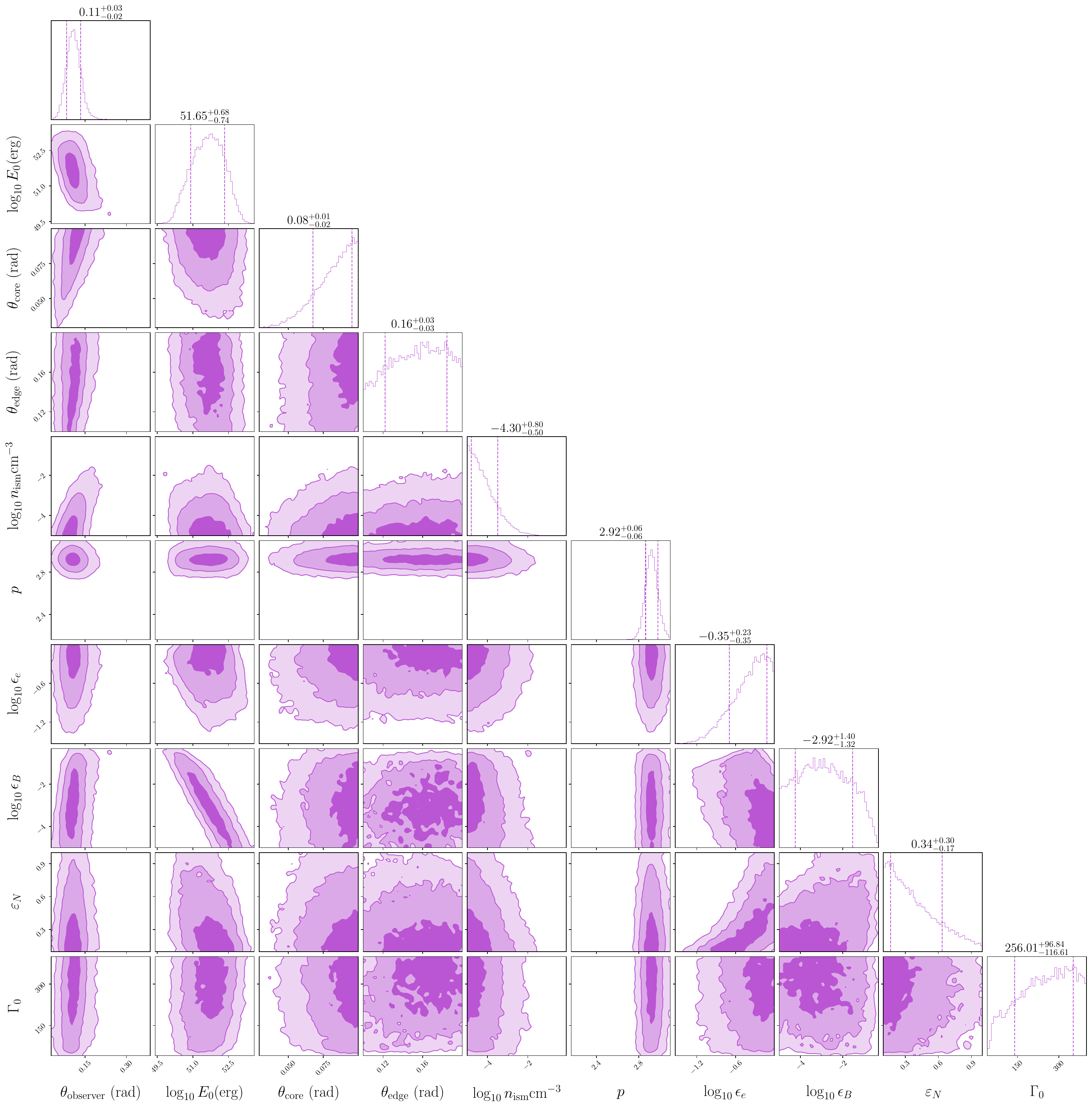}
    \caption{The posterior distribution for the model parameters used to fit the afterglow in Figure\,\ref{fig:afpy}. We excluded the optical and NIR data obtained after $t>10$~d. The model fit used {\sc nessai} as the sampler with a Gaussian likelihood via {\sc redback}. Parameter descriptions are given in Table\,\ref{tab:prior-posterior}.}
\end{figure*}

\section{Prospector modelling assuming HST second epoch data is due to host}\label{ap:prosp}
\begin{figure*}
    \centering
    \includegraphics[width=\textwidth]{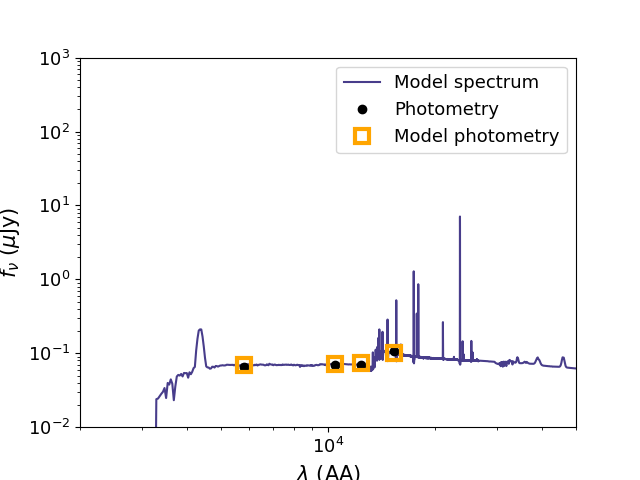}
    \caption{The best-fit galaxy model obtained with Prospector is shown. The best-fit parameters are redshift $z=3.5\pm0.7$, metallicity $\log Z/Z_\odot=-1\pm0.7$, and mass of the galaxy formed in the most-recent star formation episode $M_{\mathrm{star}} = (5^{+7}_{-4} )\times 10^9$~M$_\odot$. Currently 60\% of this mass remains. The current star formation rate is 0.7~M$_\odot$/yr. Note that there are systematic uncertainties, and we estimate that these are more important for the derived population age and metallicity than for the present day stellar mass.}
\end{figure*}

%
\bsp	
\label{lastpage}
\end{document}